\documentclass[preprint,aps,showpacs,preprint,floatfix]{revtex4-1} \usepackage{graphicx}
\usepackage{subfigure}
\usepackage{array}
\usepackage{color}
\usepackage{amsmath}
\usepackage{amsxtra}
\usepackage{tablists}
\usepackage{diagbox}
\usepackage{amstext}
\usepackage{amssymb}
\usepackage{latexsym}
\usepackage{dsfont}
\usepackage{verbatim}
\usepackage{mathrsfs}
\usepackage{graphicx}
\usepackage[position=t,singlelinecheck=off]{subfig}
\usepackage{caption}
\usepackage{multirow}
\usepackage{marvosym}
\usepackage{floatrow}
\usepackage[utf8]{inputenc}
\usepackage{caption}
\newfloatcommand{capbtabbox}{table}[][\FBwidth]

\newcommand{\beq}{\begin{equation}}
\newcommand{\eeq}{\end{equation}}
\newcommand{\bea}{\begin{eqnarray}}
\newcommand{\eea}{\end{eqnarray}}

\newcommand{\bs}[1]{{\boldsymbol{#1}}}
\newcommand{\la}{\langle}
\newcommand{\ra}{\rangle}

\newcommand{\rb}{]}

\RequirePackage[normalem]{ulem}

\begin{document}

\title{
Exploring $HVV$ amplitudes with $CP$ violation by decomposition and on-shell scattering amplitude methods}

\author{Ke-Yao Feng$^1$\footnote{fengkeyao@snnu.edu.cn}, Xia Wan$^1$\footnote{wanxia@snnu.edu.cn}, You-Kai Wang$^1$\footnote{wangyk@snnu.edu.cn}, Chao Wu$^2$\footnote{wuch7@itp.ac.cn}}
\affiliation{$^1$School of physics and Information Technology, Shaanxi Normal University,
Xi'an 710119, China}
\affiliation{$^2$ CAS Key Laboratory of Theoretical Physics, Institute of Theoretical Physics,
Chinese Academey of Sciences, 
Beijing 100190, P. R. China}

\begin{abstract}
$CP$ violation may play an important role in Baryogenesis in early universe, and should be examined at colliders comprehensively. We study $CP$ properties of $HVV$ vertexes between Higgs and gauge boson pairs with defining a $CP$ violation phase angle $\xi$, which indicates the mixture of $CP$-even and $CP$-odd Higgs states in $HVV$ in new physics. A series of $HVV$ amplitudes $H\to\gamma\gamma, H\to\gamma V\to \gamma \ell\ell$, and $H\to VV\to 4\ell$ with $CP$ phase angle are studied systematically, which explains explicitly why $CP$ violation could only be probed in $4\ell$ process independently. We get a novel amplitude decomposition relation which illustrates if two preconditions (multilinear momentum dependent vertexes and current $J_\mu$ of $V\to \ell^+ \ell^-$ is formally proportional to a photon's polarization vector) are satisfied, an high-point amplitude can be decomposed into a summation of a series of low-point amplitudes. As a practical example, the amplitude of  $H\to\gamma V\to \gamma \ell\ell$, and $H\to VV\to 4\ell$ processes can be decomposed into summation of many $H\to\gamma\gamma$ amplitudes. Meanwhile, we calculate these amplitudes in the framework of on-shell scattering amplitude method, with considering both massless and massive vector gauge bosons with $CP$ violation phase angle. The above two approaches provides consistent results and exhibit clearly the $CP$ violation $\xi$ dependence in the amplitudes.

\end{abstract}

\maketitle

\section{Introduction\label{section:introduction}}


There are two kinds of $CP$ violation sources in Standard Model~(SM),
one is weak $CP$ violation in Cabibbo-Kobayashi-Maskawa matrix~\cite{Cabibbo:1963yz,10.1143/PTP.49.652},
the other is strong $CP$ violation related to topological charge in QCD vacuum~\cite{tHooft:1976rip,tHooft:1976snw,Peccei:1977hh}.
Both of them have relations with Higgs Yukawa couplings.
The CKM matrix originates from general Yukawa coupling matrices for three generations~\cite{Chivukula:1987py,Peskin:1995ev},
the $\theta$ angle in QCD vacuum could rotate to the complex phase of mass matrix by chiral transformation~\cite{Peccei:2006as}.
Even though, the SM Higgs boson is a $CP$-even scalar with $CP$-conserving interactions.
By contrast, in new physics beyond SM (BSM), $CP$ violation usually relates to Higgs bosons.
 One reason is there often exist scalars and pseudoscalars instead of
one single scalar in SM. A mixture of scalar and pseudoscalar is natural.
Such as in Two Higgs Doublet Model (THDM)~\cite{Branco:2011iw},
Minimal Supersymmetric Standard model (MSSM)~\cite{Djouadi:2005gj}, and Composite Higgs Model~\cite{Panico:2015jxa},
pseudoscalar always appears and there are no simple rule to forbid a mixture between the scalar and pseudoscalar.
Except for theoretical naturalness and generality,
one practical motivation for new $CP$ violation source comes from
the matter-antimatter asymmetry
observed in our Universe~\cite{Steigman:1976ev,Steigman:2008ap,Cohen:1987vi}.
In electroweak Baryogenesis mechanism~\cite{Kuzmin:1985mm},
$CP$ violation
plus sphaleron transition~\cite{Klinkhamer:1984di} could produce baryon and lepton
number violation during electroweak phase transition, but
the $CP$ violation ratio in SM is too small to fulfil the
quantity of the observed matter-antimatter asymmetry~\cite{Gavela:1993ts,Gavela:1994dt,Huet:1994jb}.
Therefore
new $CP$ violation source must be added to obtain electroweak Baryogenesis.

We choose
 two model-independent framework to study $CP$ violation: one is a traditional way through 
 additional effective Lagrangian terms,
 the other is using on-shell scattering amplitude method to analyze amplitudes.

Adding new $CP$-violating terms in the Lagrangian is an convenient, effective description of the new couplings beyond the SM. The new terms can be $CP$ conserved or $CP$ violated, but should obey the Lorentz and gauge  invariance. For a specific new physics model, its new Higgs couplings can be simplified into this effective Lagrangian terms when other couplings are small enough to be omitted. Therefore, constraints on these new Higgs couplings in effective Lagrangian provide concrete limitations for model buildings with certain gauge symmetries.


On-shell method is a novel tool to deal with amplitudes directly, even with no
Lagrangian and Feynman diagram needed~\cite{Elvang:2015rqa}. It starts from on-shell
particle states instead of field,
sets up constraints,
exploits analytical properties such as poles and branch cuts,
 then gets an available amplitude.
Specifically,
a $3$-point massless (or 1 massive 2 massless) amplitude could be fixed
 by locality and little group scaling~\cite{Elvang:2015rqa,Cheung:2017pzi,Arkani-Hamed:2017jhn},
then a $n+1$-point tree amplitude could be constructed from $n$-point amplitudes
through recursion relations.
In this way all tree amplitudes could be obtained and they have clear mathematical structures.

We focus on $H\to \gamma \gamma$, $H\to \gamma \ell \ell$ and
$H\to 4\ell$ processes to analyze their BSM amplitudes. 
At Large Hadron collider (LHC), the $H\to \gamma \gamma$
and $H\to ZZ \to 4\ell$ processes are Higgs discovery channel~\cite{CMS:2012qbp,ATLAS:2012yve},
which have the advantage of clean background and relative large signal.
They are also golden channels for precise measurement of Higgs properties
~\cite{Campbell:2013una,Caola:2013yja,Dixon:2013haa,CMS:2019ekd}.
In our previous research, we notice that $CP$ violation phase
could not be probed solely in $H\to \gamma \gamma$ or $H\to \gamma \ell \ell$ processes
without interference from background~\cite{Wan:2017qiq,Chen:2017plj}. By contrast,
in $H\to ZZ \to 4\ell$ processes $CP$ violation could be probed lonely through
its kinematic angles~\cite{He:2019kgh,CMS:2019ekd,Chatrchyan:2013mxa,Anderson:2013afp}. These could be explained clearly
at amplitude level after we get a compact formula. 
We explore the relations between these BSM amplitudes through the above two independent ways in this paper. 
A decomposition relation between these amplitudes is illustrated in an interesting diagrammatic way. 
Then we calculate the same amplitude from the on-shell method (BCFW recursion relation), which can be regarded as a parallel proof of the decomposition relation. 
 Meanwhile, the massive spinor formalism are applied to prove it is also suitable for massive cases.

The outline of this paper is as follows. In section~\ref{section:SM}, we show the amplitudes of SM $HVV$ processes both at proton-proton collider and $e^+e^-$ collider.   In section~\ref{section:BSMamp}, we calculate BSM amplitudes in effective Lagrangian description.
The $H\to \gamma \gamma$, $H\to \gamma \ell \ell$ and
$H\to 4\ell$ processes correspond separately to  $3$, $4$, $5$-point amplitudes.
In section~\ref{section:decomposition}, we deduce decomposition relations for these amplitudes.
In section~\ref{section:on-shell}, we reproduce these BSM amplitudes in on-shell scattering amplitude approach.
In section~\ref{section:on-shellmassive}, the BSM amplitudes in on-shell scattering amplitude are generalized to massive spinor cases.
Section~\ref{section:summary} is  summary and discussion.

\section{SM HVV Helicity amplitudes\label{section:SM}}

Experimentally, SM/BSM $HVV$ couplings can be measured/exploited at the existed proton-proton collider such as the LHC or the future $e^+e^-$ collider such as 
the Circular Electron Positron Collider 
(CEPC), the International Linear Collider (ILC) and the Compact Linear Collider (CLIC), etc.~. To study the BSM $HVV$ couplings with $CP$ violation, its interference with the corresponding SM process (include or not include Higgs) may become the dominant contribution as generally the BSM couplings are assumed to be suppressed compared to the corresponding SM couplings. Before studying the BSM amplitudes, the amplitudes of SM $HVV$ process and the main background process are
introduced to show a global view how amplitudes works for physics process at colliders. Analyzing amplitudes could unveil 
some mysteries that are not clear at the observable level.
As followings we take specific $HZZ$ related processes as an example, at proton-proton collider and $e^+e^-$ collider separately. 
For the example at proton-proton collider, we focus on process-dependent 
amplitudes, show 
amplitudes of signal process and background process and discuss how these amplitudes are used for experimental predictions. For the example at 
$e^+e^-$ collider, we focus on process-independent amplitudes,
that is, the amplitudes with all external particles outgoing, which relate to process-dependent 
amplitudes by crossing symmetry.

\subsection{At proton-proton collider}

\begin{figure}[!htbp]
\begin{minipage}{0.45\linewidth}
\centerline{\includegraphics[width=1.0\textwidth]{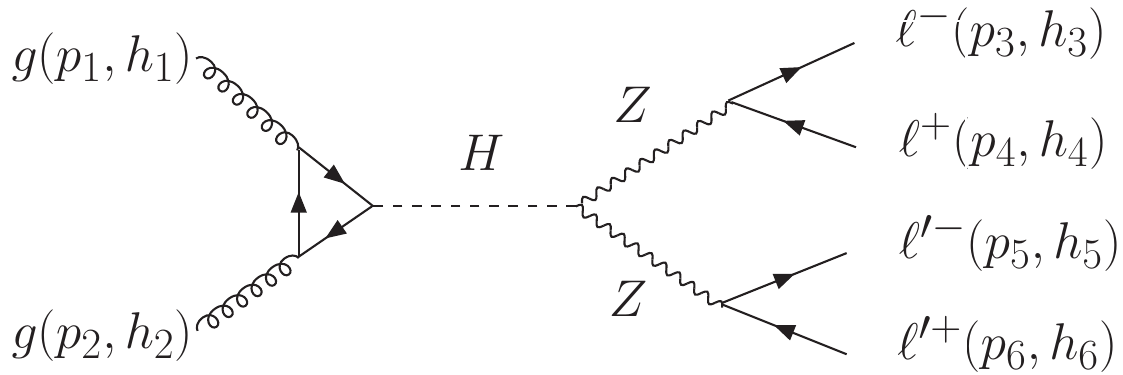} }
\end{minipage}
\hspace{1cm}
\begin{minipage}{0.45\linewidth}
\centerline{\includegraphics[width=1.0\textwidth]{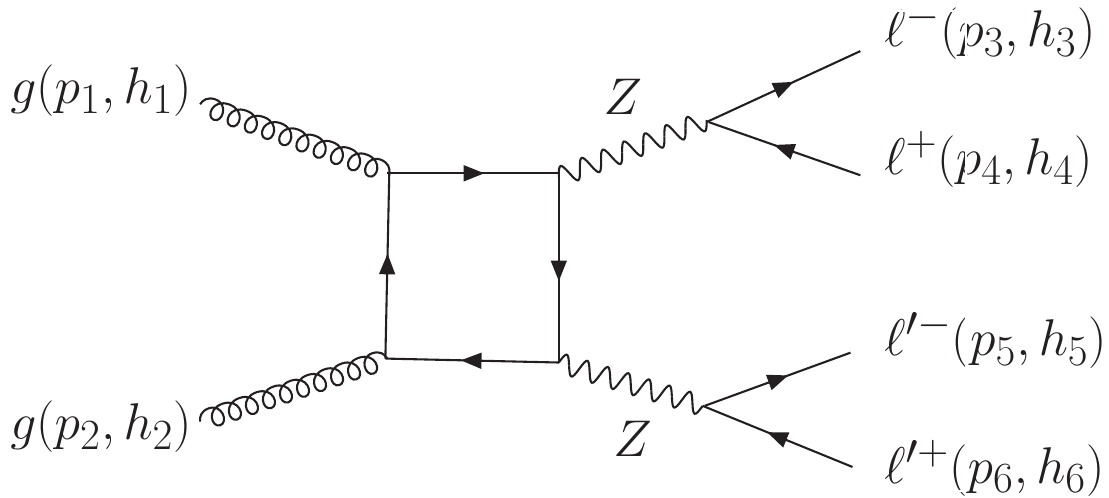} }
\end{minipage}\\
\begin{minipage}{0.45\linewidth}
(a)
\end{minipage}
\hspace{1cm}
\begin{minipage}{0.45\linewidth}
(b)
\end{minipage}
\caption{\it
Feynman diagrams of SM $HZZ$ related process and its main background with 4$\ell$ final states at the LHC. }
\label{SMHtoZZ}
\end{figure}

The $gg\to H \to ZZ \to 4 \ell$ process at LHC is sensitive to BSM $HVV$ couplings. We draw the Feynman diagrams of  this signal channel and its main background as in Fig.~\ref{SMHtoZZ}.
In Fig.~\ref{SMHtoZZ}(a) the Higgs production process $gg\to H$ are mediated by the top quark loop. Fig.~\ref{SMHtoZZ}(b) represents $gg\to ZZ \to 4 \ell$ box process without Higgs, which is important in the off-shell Higgs region.
Studying the interference between the signal 
and this continuum background in the off-shell Higgs region could give a stringent bound on Higgs width~\cite{CMS:2014quz, CMS:2022ley}, and also BSM $HZZ$ couplings~\cite{CMS:2022ley,He:2019kgh}.
In Fig.~\ref{SMHtoZZ} the $\ell$ and $\ell^\prime$ have different flavors. 
If we study the process of $4\ell$ with same flavors, two more diagrams which describe 
another pairing of $4\ell$ should be added. Nevertheless, their amplitudes are similar~\cite{He:2019kgh}.

The triangle top quark loop in $gg\to H $ could be integrated and described by an effective $ggH$ coupling such that the helicity amplitudes of $ggH$ are shown as~\cite{Chen:2017plj}

\bea
 \mathcal{M}^{gg\to H}(1^{+}_g,2^{+}_g)&=&\frac{2c_g}{v}[12]^2~, \nonumber\\
 \mathcal{M}^{gg\to H}(1^{-}_g,2^{-}_g)&=&\frac{2c_g}{v}\langle12\rangle^2~.
 \label{eqn:ggh}
\eea
with 
\beq
\frac{c_g}{v}=\frac{1}{2}\sum_f\frac{\delta^{a b}}{2}\frac{i}{16\pi^2}g^2_s4e
\frac{m_f^2}{2M_W s_W}\frac{1}{M^2_H}(2+s_{12}(1-\tau_H)C^{\gamma\gamma}_0(m_f^2))~,
\label{eqn::FggH}
\eeq
where $v=246$ GeV is the vacuum expectation value of the Higgs,  $a, b=1,...,8$ are $SU(3)_c$ adjoint representation indices for the gluons,
$\tau_H=4m_f^2/M^2_{H}$, and the $C^{\gamma\gamma}_{0}(m^2)$ function is Passarino-Veltman three-point scalar functions \cite{Passarino:1978jh}.
The $\langle  i j \rangle$ and $[ ij ]$ are followed as the conventions in Ref.~\cite{Dixon:1996wi,Dixon:2013uaa}:
\bea
&&\langle ij \rangle \equiv \langle i^-|j^+\rangle= \overline{u_-(p_i)} u_+(p_j),\quad[ ij ]\equiv \langle i^+|j^-\rangle = \overline{u_+(p_i)} u_-(p_j), \nonumber\\
&&\langle ij \rangle[ ji ] = 2 p_i \cdot p_j,\quad s_{ij} = (p_i+p_j)^2,\quad
\epsilon_\mu^{\pm}(p_i,q)=\pm\frac{\langle q^{\mp}|\gamma_{\mu}|p_i^{\mp}\rangle}{\sqrt{2}\langle q^{\mp}|p_i^{\pm}\rangle},
\label{eq:epsilon}
\eea
where $p_i$ are momentum of external legs,
$q$ is the reference momentum that reflect the freedom of gauge transformation, $\epsilon^{\pm}(p_i,q)$ is for outgoing photons with $\pm$ helicities.
Notice that the gluons are incoming in Eq.~\eqref{eqn:ggh}, if 
let them outgoing, 
the amplitudes just need an exchange 
between $\langle\rangle$ and $[]$
because of the crossing symmetry.

The helicity amplitudes of $H\to ZZ \to 4\ell$ are \cite{He:2019kgh}
\bea 
\mathcal{M}^{H\to ZZ \to 4\ell}(3^{-}_{\ell^- },4^{+}_{\ell^+},5^{-}_{\ell^- },6^{+}_{\ell^+})
&=&f\times l_e^2 \frac{M_W^2}{\cos^2{\theta_W}}\langle 35\rangle[46],\nonumber \\
\mathcal{M}^{H\to ZZ \to 4\ell}(3^{-}_{\ell^- },4^{+}_{\ell^+},5^{+}_{\ell^- },6^{-}_{\ell^+})
&=&f\times l_e r_e \frac{M_W^2}{\cos^2{\theta_W}}\langle 36\rangle[45],\nonumber \\
\mathcal{M}^{H\to ZZ \to 4\ell}(3^{+}_{\ell^- },4^{-}_{\ell^+},5^{-}_{\ell^- },6^{+}_{\ell^+})
&=&f\times l_e r_e \frac{M_W^2}{\cos^2{\theta_W}}\langle 45\rangle[36],\nonumber \\
\mathcal{M}^{H\to ZZ \to 4\ell}(3^{+}_{\ell^- },4^{-}_{\ell^+},5^{+}_{\ell^- },6^{-}_{\ell^+})
&=&f\times r_e^2 \frac{M_W^2}{\cos^2{\theta_W}}\langle 46\rangle[35],
\label{H4lamp}
\eea
where the common factor $f$ is defined as 
\beq
f=-2ie^3\frac{1}{M_W\sin\theta_W} P_Z(s_{34}) P_Z(s_{56}),
\eeq
 with 
 \beq
  P_X(s)=\frac{1}{s-M^2_X+iM_X\Gamma_X}
  \label{propagator}
 \eeq
are the propagator of particle $X$.
So the total amplitude of $gg \to H \to ZZ \to 4\ell$ is
\beq
\mathcal{M}^{gg\to H\to ZZ \to 4\ell}=
\mathcal{M}^{gg\to H}\times P_H(s_{12}) \times \mathcal{M}^{H\to ZZ \to 4\ell }~.
\label{LHCamp}
\eeq

The amplitude of the box process $gg \to ZZ \to 4\ell$ in Fig.\ref{SMHtoZZ}(b) are complicated due to the box loop integration. Its full analytical form could be found in Ref.~\cite{Campbell:2013una}, which is coded in MCFM package~\cite{Campbell:2019dru}. 
Meanwhile we could do phase space integral and get numerical cross sections in MCFM package.

When studying the observable effects of the BSM $HVV$ couplings, we could add BSM amplitudes to MCFM package
and get total and partial cross section results.
As the analytical form of each amplitude
is clear, each contribution for the 
cross section can be singly shown.
For example,
the interference contribution 
from SM Higgs process and BSM Higgs process can be calculated by selecting ${\text {Re}}(\mathcal{M}^{SM}_H \mathcal{M}^{\ast{BSM}}_{H})$ part in the code.
In Higgs off-shell region, the interference 
contribution between continuum 
background and BSM process is also important, so the 
 ${\mbox {Re}}(\mathcal{M}^{SM}_{\text {box}} \mathcal{M}^{\ast{BSM}}_{H})$ needs to be singly focused on. More details could be found in Ref.~\cite{He:2019kgh}.

\subsection{At $e^+e^-$ Collider}

\begin{figure}[!htbp]
\begin{minipage}{0.45\linewidth}
\centerline{\includegraphics[width=1.0\textwidth]{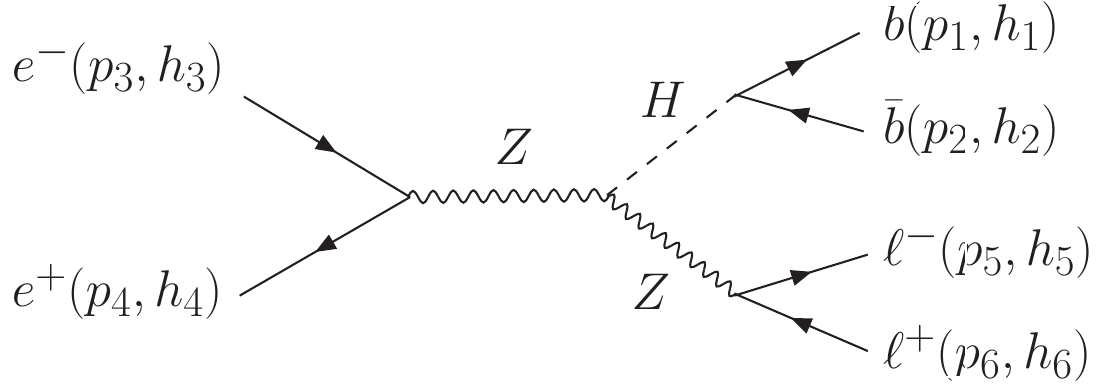} }
\end{minipage}
\hspace{1cm}
\begin{minipage}{0.45\linewidth}
\centerline{\includegraphics[width=1.0\textwidth]{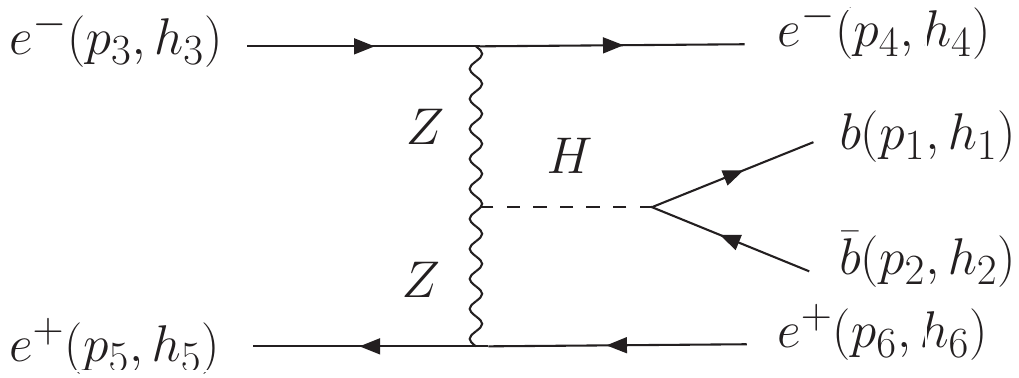} }
\end{minipage}\\
\begin{minipage}{0.45\linewidth}
(a)
\end{minipage}
\hspace{1cm}
\begin{minipage}{0.45\linewidth}
(b)
\end{minipage}
\caption{\it
Feynman diagrams of SM $HZZ$ related process at the CEPC.
}
\label{SMHtoZZCEPC}
\end{figure}

At $e^+e^-$ Collider, two main processes
with $HZZ$ coupling are $ZH$ process and vector boson fusion process.
Their Feynman diagrams are shown as in Fig.~\ref{SMHtoZZCEPC}.
In these two processes the $e^+e^-$ are incoming particles and $b\bar{b} \ell^+\ell^- $ (or $b\bar{b} e^+e^-$) are outgoing particles.
As crossing symmetry illustrates that
an incoming particle could be replaced by an outgoing antiparticle and leave the S-matrix unchanged, we could calculate 
the amplitude with all particles outgoing firstly and then deduce 
the amplitude for physics process just by relabeling momenta, helicities and particle properties. We won't go to details about using crossing symmetry since it is actually trivial once rules are set, but focus on how to write the process-independent amplitudes with all external particles outgoing. Later we also keep this convention for all BSM amplitudes.
So the amplitudes we need are same for Fig.~\ref{SMHtoZZCEPC}(a) 
and Fig.~\ref{SMHtoZZCEPC}(b) if $\ell$ is assumed to be $e$ or both them are assumed to be massless. They could be written as
\beq
 \mathcal{M}({1_b, 2_{\bar{b}}, 3_{e^-},4_{e^+},5_{l^-},6_{l^+}})
 = \mathcal{M}(1_b, 2_{\bar{b}}, I_H)\times 
 P_H(s_{12}) \times \mathcal{M}(I^\prime_H, 3_{e^-},4_{e^+},5_{l^-},6_{l^+} )~,
 \label{CEPCamp}
 \eeq
 where $I_H$, $I^\prime_H$ represents the 
mediate $H$ which are broken into two parts and appear in each smaller amplitudes. 
The amplitude $\mathcal{M}(I^\prime_H, 3_{e^-},4_{e^+},5_{l^-},6_{l^+} )$ has been 
calculated in Eq.~\eqref{H4lamp} except flipping the momentum of the Higgs boson to be outgoing. 
Actually, because Higgs is a scalar, the amplitude will remain unchanged in this case.   
One new amplitude we should pay attention is 
$\mathcal{M}(1_b, 2_{\bar{b}}, I_H)$, which is

\bea
\mathcal{M}(2^-_b,3^-_{\bar{b}})&=&\frac{-i m_b}{v}\langle12\rangle,\nonumber \\
\mathcal{M}(2^+_b,3^+_{\bar{b}})&=&\frac{-i m_b}{v}[12].
\label{Hbb}
\eea
The external $b$,$\bar{b}$ quarks are assumed to be massless in high energy limit. The $H b \bar{b}$ coupling is still fixed to be proportional to 
$m_b$. The amplitudes for massive particles
are realized in massive spinor formalism and 
need to use little group indices~\cite{Arkani-Hamed:2017jhn,Durieux:2019eor,Wu:2021nmq}, which relates $\langle12\rangle$ 
to bolded $\langle \bs{12}\rangle$ for example.
We study them later in Section~\ref{section:on-shellmassive}.

\section{BSM $HVV$ helicity amplitudes\label{section:BSMamp}}

In this section firstly we 
introduce the BSM $HVV$ effective couplings
 and define the $CP$ violation phase.
Then we calculate their amplitudes. At last we discuss the interference contribution from the BSM amplitudes.

\subsection{$HVV$ effective couplings}

In SMEFT~\cite{Buchmuller:1985jz, Grzadkowski:2010es,Brivio:2017vri},
the complete form of higher-dimensional operators can be written as
\beq
\mathcal{L}=\mathcal{L}_{\rm SM}+\frac{1}{\Lambda}\sum_k C_k^{5}\mathcal{O}_k^{5}
+\frac{1}{\Lambda^2}\sum_k C_k^{6}\mathcal{O}_k^{6}+\mathcal{O}(\frac{1}{\Lambda^3})~,
\label{complete_form}
\eeq
where $\Lambda$ is energy scale of new physics, and $C_k^{i}$ with $i=5,6$
are Wilson loop coefficients.

BSM $HVV$ ($V$ represents $\gamma$, $Z/W$ boson) vertices start from dimension-six operators $\mathcal{O}_k^{6}$.
In Warsaw basis~\cite{Grzadkowski:2010es}, they are
\bea
&&\mathcal{O}^6_{\Phi D} =(\Phi^{\dagger}D^{\mu}\Phi)^{\ast}(\Phi^{\dagger}D^{\mu}\Phi), \nonumber \\
&&\mathcal{O}^6_{\Phi W}=
\Phi^\dagger \Phi W^{I}_{\mu\nu}W^{I\mu\nu},~~
\mathcal{O}^6_{\Phi B}= \Phi^\dagger\Phi B_{\mu\nu}B^{\mu\nu},~~
\mathcal{O}^6_{\Phi WB}= \Phi^\dagger \tau^I \Phi W^{I}_{\mu\nu}B^{\mu\nu},
\nonumber \\
&& \mathcal{O}^6_{\Phi \tilde{W}}= \Phi^\dagger\Phi \tilde{W}^{I}_{\mu\nu}W^{I\mu\nu},
~~\mathcal{O}^6_{\Phi \tilde{B}}= \Phi^\dagger\Phi \tilde{B}_{\mu\nu}B^{\mu\nu},
~~\mathcal{O}^6_{\Phi \tilde{W}B}= \Phi^\dagger \tau^I \Phi \tilde{W}^{I}_{\mu\nu}B^{\mu\nu},
\label{operatorphiw}
\eea
where $\Phi$ is a doublet representation under the $SU(2)_L$ group and the aforementioned Higgs field $H$ is one of its four components;
$D_\mu=\partial_\mu-i g W^{I}_{\mu}T^{I}-ig^{\prime}YB_\mu$, where $g$ and $g^\prime$ are coupling constants, $T^{I}=\tau^{I}/2$, where $\tau^{I}$ are Pauli matrices, $Y$ is the $U(1)_Y$ generator;
$W^{I}_{\mu\nu}=\partial_\mu W^{I}_{\nu}-\partial_\nu W^{I}_\mu-g\epsilon^{IJK}W^{J}_\mu W^{K}_\nu$,
$B_{\mu\nu}=\partial_\mu B_\nu-\partial_\nu B_\mu$,
$\tilde{X}_{\mu\nu}= \frac{1}{2}\epsilon_{\mu\nu\rho\sigma}X^
{\rho\sigma}$,

After spontaneous symmetry breaking, we get $HVV$ effective interactions,
\beq
\mathcal{L}^{int}=-\frac{c_{VV}}{v}HV^{\mu\nu}V_{\mu\nu}
-\frac{\tilde{c}_{VV}}{v}HV^{\mu\nu}\tilde{V}_{\mu\nu}~,
\label{L_int}
\eeq
$c_{VV}$, $\tilde{c}_{VV}$ are real numbers that originate
from Wilson loop coefficients,
$V$ represents vector boson.
A detailed formula about $c_{VV}$, $\tilde{c}_{VV}$ and Wilson loop coefficients
$C_k^6$ could be found in Ref.~\cite{He:2019kgh}. A standard analysis based on SMEFT should be a global study involving all dimension-6 operators. Here we concentrate only on the new $HVV$ terms. 

 For the origin of the dimension-6 operators, they can come from the loop momentum integration in the loop diagrams with multi-outlegs. The virtual particles in the loop can be both SM particles and BSM new particles. The difference between the two cases is the SM processes have definite dimension-6 coupling coefficients while the dimension-6 coefficients in New Physics(NP) are still to be determined. 

The $CP$ violation phase could be defined as
\beq
\xi~\equiv~ {\rm tan}^{-1}(\tilde{c}_{VV}/c_{VV})~,
~~~~~~~~{\rm when~~Arg}(\tilde{c}_{VV}/c_{VV})=0~~{\rm or}~~\pi~,
\label{xi}
\eeq
where $\xi=0~(\frac{\pi}{2})$ represents a pure $CP$-even~(-odd) $HVV$ vertex.
$\xi\neq 0$ means $CP$ violation and
$\xi=\frac{\pi}{2}$ corresponds to maximal $CP$ violation
if other Higgs vertices are supposed to be $CP$-even.
In amplitudes we will see that $\xi$ appears as a phase,
which changes sign under $CP$ transformation. That is why we name it as
$CP$ violation phase.
Meanwhile,
\beq
c^S_{VV}\equiv\sqrt{c^2_{VV}+\tilde{c}^2_{VV}}
\label{cs}
\eeq
could be defined as the amplitude moduli,
which is proportional to signal strength in collider experiment.

\subsection {Helicity amplitudes}

In following sections, for simplification, we only take the amplitude of Higgs decay with BSM $HVV$ vertex as an example to illustrate the decomposition relation. It is also the process-independent amplitude since Higgs boson is a scalar and the amplitude is free of its incoming or outgoing. 
Full amplitudes with Higgs production and decay, can be easily obtained by multiplying the BSM Higgs decay amplitudes with the partial amplitude of Higgs production $\mathcal{M}^{gg\to H}$ in Eq.~(\ref{eqn:ggh}) and a Higgs propagator in Eq.~(\ref{propagator}) at the proton-proton collider, or by multiplying with the partial amplitude of $H\to b\bar{b}$ Eq.~(\ref{Hbb}) and a Higgs propagator in Eq.~(\ref{propagator}) at the $e^+e^-$ collider, similar as Eq.~(\ref{LHCamp}) or Eq.~\eqref{CEPCamp} as discussed in section~\ref{section:SM}.

\begin{figure}[!htbp]
\begin{minipage}{0.3\linewidth}
\centerline{\includegraphics[width=1.0\textwidth]{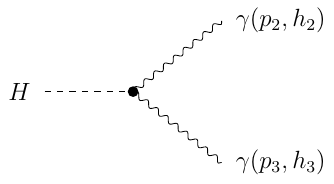} }
\end{minipage}
\begin{minipage}{0.3\linewidth}
\centerline{\includegraphics[width=1.0\textwidth]{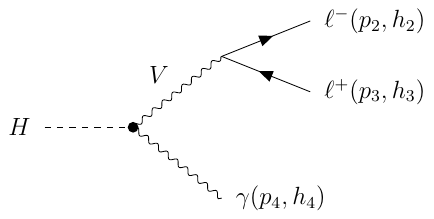} }
\end{minipage}
\begin{minipage}{0.3\linewidth}
\centerline{\includegraphics[width=1.0\textwidth]{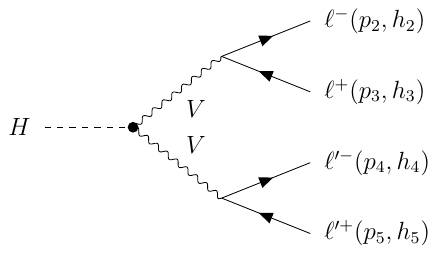} }
\end{minipage}

\caption{\it
Feynman diagrams of $H\to \gamma \gamma$, $H\to V \gamma\to  \ell \ell \gamma$ and $H\to V V \to  2\ell 2\ell^\prime$ from left to right.
Each $HVV$ vertex is dotted as an effective coupling.
}
\label{fig:Feynman-diagrams}
\end{figure}

Feynman diagrams with effective $HVV$ couplings are shown in Fig.~\ref{fig:Feynman-diagrams}.
After some calculations, the helicity amplitudes are as follows.
\begin{itemize}
\item For process $H\to\gamma \gamma$,
 \bea
\mathcal{M}(2^{+}_\gamma,3^{+}_\gamma)
&=&\frac{2c^S_{\gamma\gamma}}{v}e^{i\xi}[23]^2~, \nonumber\\
 \mathcal{M}(2^{-}_\gamma,3^{-}_\gamma)
&=&\frac{2c^S_{\gamma\gamma}}{v}e^{-i\xi}\langle23\rangle^2~,\nonumber\\
 \mathcal{M}(2^{+}_\gamma,3^{-}_\gamma)&=&0~, \nonumber\\
 \mathcal{M}(2^{-}_\gamma,3^{+}_\gamma)&=&0~,
\label{eqn:hgammagamma}
\eea
where we use $\mathcal{M}(2_\gamma^{h_2},3_\gamma^{h_3})$ to represent
$\mathcal{M}(1_H^{h_1},2_\gamma^{h_2},3_\gamma^{h_3})$
since $h_1$ is trivially zero for all cases,
$h_i$s are helicities of external legs with momentum outgoing.
The results show that helicities of the two photons should keep same sign
because the spin of Higgs is zero and total angular momenta conserves.
Under $CP$ transformation
$\mathcal{M}(2^{+}_\gamma,3^{+}_\gamma)$ changes to
$\mathcal{M}(2^{-}_\gamma,3^{-}_\gamma)$. Analytically it corresponds that
$\langle i j \rangle$ changes to $[i j]$. Thus in Eq.~\eqref{eqn:hgammagamma},
a general nonzero $\xi$ represents $CP$ violation.

\item For process $H\to V \gamma\to  \ell \ell \gamma$,
\bea
\mathcal{M}(2^{-}_{\ell^- },3^{+}_{\ell^+},4^{-}_\gamma)
&=&f^-_{V}(s_{23})\times \frac{2c^S_{\gamma V}}{v}e^{-i\xi}[23]\langle 24\rangle^2,
\nonumber\\
\mathcal{M}(2^{-}_{\ell^- },3^{+}_{\ell^+},4^{+}_\gamma)
&=&f^-_{V}(s_{23})\times \frac{2c^S_{\gamma V}}{v}e^{i\xi}\langle 23\rangle[34]^2,
\nonumber\\
\mathcal{M}(2^{+}_{\ell^- },3^{-}_{\ell^+},4^{+}_\gamma)
&=&f^+_{V}(s_{23})\times \frac{2c^S_{\gamma V}}{v}e^{i\xi}\langle 23\rangle[24]^2,
\nonumber\\
\mathcal{M}(2^{+}_{\ell^- },3^{-}_{\ell^+},4^{-}_\gamma)
&=&f^+_{V}(s_{23})\times \frac{2c^S_{\gamma V}}{v}e^{-i\xi}[23]\langle 34\rangle^2,
\label{eqn:hzgamma}
\eea
where $s_{23}=(p_2+p_3)^2$, $f^-_{V}(s)=\sqrt{2}~e~l_VP_V(s)$
and $f^+_{V}(s)=-\sqrt{2}~e~r_VP_V(s)$,
$P_V(s)=\frac{1}{s-M^2_V}$ is the propagator of the gauge boson,
$l_V$ and $r_V$ are the left-handed and right-handed couplings between vector boson and leptons, and leptons are supposed to be massless.
The remaining helicity amplitudes are equal to zero and thus not listed.

\item For process $H\to V V \to  2\ell 2\ell^\prime$,
{\small
\bea
\mathcal{M}(2^{-}_{\ell^-},3^{+}_{\ell^+ },4^{-}_{\ell^{\prime -}},5^{+}_{\ell^{\prime +}})
=f^{-}_{V}(s_{23})f^{-}_{V}(s_{45}) \frac{2c^S_{VV}}{v}
\left( e^{i\xi}\langle 23\rangle \langle 45\rangle [35]^2+e^{-i\xi}
[23][45]\langle 24\rangle^2 \right),
\nonumber\\
\mathcal{M}(2^{-}_{\ell^-},3^{+}_{\ell^+ },4^{+}_{\ell^{\prime -}},5^{-}_{\ell^{\prime +}})
=f^{-}_{V}(s_{23})f^{+}_{V}(s_{45})  \frac{2c^S_{VV}}{v}
\left( e^{i\xi}\langle 23\rangle \langle 45\rangle [34]^2+e^{-i\xi}
[23][45]\langle 25\rangle^2 \right),
\nonumber\\
\mathcal{M}(2^{+}_{\ell^-},3^{-}_{\ell^+ },4^{-}_{\ell^{\prime -}},5^{+}_{\ell^{\prime +}})
=f^{+}_{V}(s_{23})f^{-}_{V}(s_{45}) \frac{2c^S_{VV}}{v}
\left( e^{i\xi}\langle 23\rangle \langle 45\rangle [25]^2+e^{-i\xi}
[23][45]\langle 34\rangle^2 \right),
\nonumber\\
\mathcal{M}(2^{+}_{\ell^-},3^{-}_{\ell^+ },4^{+}_{\ell^{\prime -}},5^{-}_{\ell^{\prime +}})
=f^{+}_{V}(s_{23})f^{+}_{V}(s_{45})  \frac{2c^S_{VV}}{v}
\left( e^{i\xi}\langle 23\rangle \langle 45\rangle [24]^2+e^{-i\xi}
[23][45]\langle 35\rangle^2 \right),
\label{eqn:hzz}
\eea
where $VV$ could be $\gamma\gamma$, or $ZZ$, or $\gamma Z$, or $W^+W^-$.
While when it represents $\gamma Z$ or $W^+W^-$, the original Lagrangian in Eq.~\eqref{L_int}
should be scaled by a factor of $2$ on the whole
to make the formula consistent.
The remaining helicity amplitudes are equal to zero.}

\end{itemize}

\subsection{Interference contribution}
By using the compact form of the amplitudes and the definition of $CP$ violation phases angle, it is interesting to compare the SM $HVV$ amplitudes with BSM $HVV$ amplitudes, and then show how to extract these BSM contributions in collider experiments.

Firstly we compare the SM $H\gamma\gamma$ amplitudes 
with BSM $H\gamma\gamma$ amplitudes. 
The SM $H\gamma\gamma$ amplitudes 
can be obtained by replacing the 
coefficient $C_g$ with  $C_\gamma$
in Eq.~(\ref{eqn:ggh}), where $C_\gamma$ represents the triangle loop integral from both top quark loop and $W$ boson loop~\cite{Djouadi:2005gi}.
The BSM $H\gamma\gamma$ amplitude are shown in Eq.~(\ref{eqn:hgammagamma}). Comparing Eq.~(\ref{eqn:ggh}) with Eq.~(\ref{eqn:hgammagamma}),
their spinor structures are the same while the $CP$ violation phase and coefficients are different. Therefore, for interference between these two amplitudes, the kinematic observables ( such as the shape of the angular distribution of the external particles) will remain unchanged except for an overall scale factor.

Next we compare the SM and BSM amplitudes in .~(\ref{H4lamp}) and Eq.(\ref{eqn:hzz}) of  $H\to ZZ \to 4\ell$ processes. Their spinor structures are completely different, as each SM amplitude has two brackets (including both $\langle\rangle$ and $[]$ ) and one term, while the BSM amplitude have four brackets and two summed terms. The two more brackets in BSM amplitudes origin from the partial derivatives in dim-6 operators as shown in Eq.~(\ref{L_int}). Therefore, the extra momentum dependence of the  BSM scattering amplitudes can be regarded as an indication of the momentum dependence of the BSM couplings. It is obvious that this interference contribution between BSM and SM amplitudes, which is proportional to the momentum of the external particles, will be enhanced in the high energy region. In other words, the interference effects are expected to 
be searched sensitively in the off-shell Higgs High energy region~\cite{He:2019kgh}.

\section{Decomposition of helicity amplitudes~\label{section:decomposition}}

Amplitudes of $CP$ violation $HVV$ processes in Eq.~\eqref{eqn:hgammagamma}\eqref{eqn:hzgamma} and \eqref{eqn:hzz}
have similar structures.
In $H\to\gamma\gamma$ and $H\to \ell\ell \gamma$ processes,
there is only one term for each helicity amplitude.
$CP$ violation phase shows as a global phase.
However, in $H\to 4\ell$ process,
two terms appear and $CP$ violation phases have reverse signs.
To explore how amplitudes change when external legs increase,
we find a decomposition relation for a particular type of $n$-particle effective interactions. Then we apply it to the $HVV$ effective interactions and derive the corresponding amplitudes.

\subsection{Proof}

\begin{figure}[!htbp]
    \begin{minipage}{0.4\linewidth}
        \centerline{\includegraphics[width=1.0\textwidth]{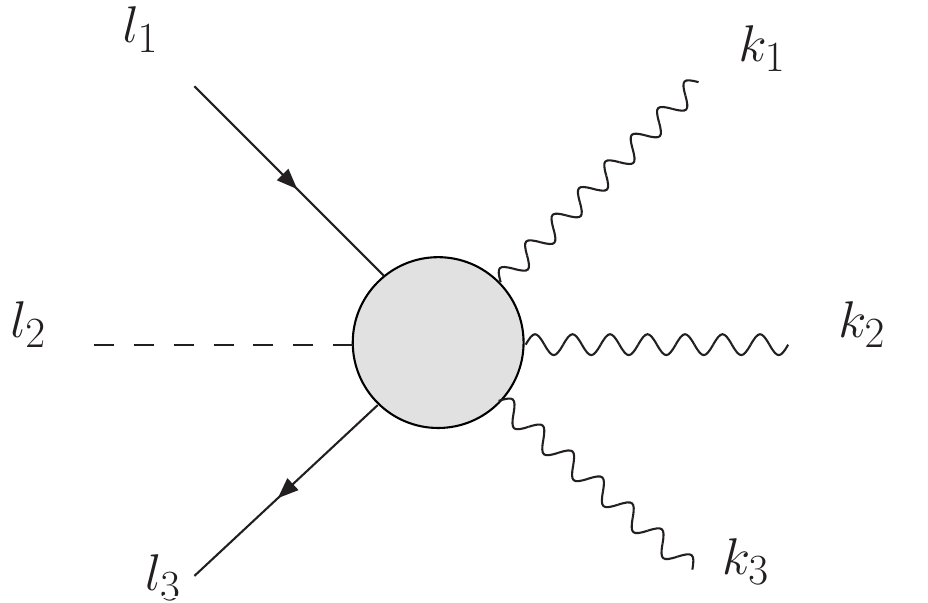} }
    \end{minipage}\qquad
    \begin{minipage}{0.4\linewidth}
        \centerline{\includegraphics[width=1.0\textwidth]{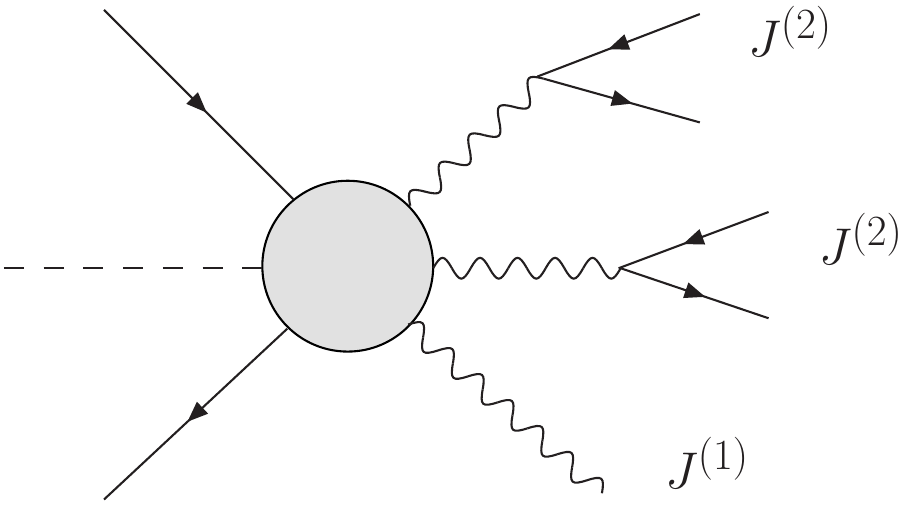} }
    \end{minipage}
    \begin{minipage}{0.4\linewidth}
        (a)
    \end{minipage}\qquad
    \begin{minipage}{0.4\linewidth}
        (b)
    \end{minipage}
    \caption{\it \centering Diagram (a) and (b) are two examples of $M_{\mathrm{lower}}$ and $M_{\mathrm{higher}}$, where the blob represents the same effective interaction. In $M_{\mathrm{lower}}$, $k_i$ and $l_i$ characterize vector bosons and other particles. In $M_{\mathrm{higher}}$, some external vector bosons are replaced with the current $J^{(2)}$ while the others are noted by $J^{(1)}$.}
    \label{fig10}
\end{figure}

Consider the amplitudes $M_{\mathrm{lower}}$ and $M_{\mathrm{higher}}$ with $m$ and more than $m$ external lines, which both include an $m$-particle effective interactions, e.g. Fig.\ref{fig10}. In $M_{\mathrm{higher}}$, all propagators are vector bosons and they are attached to the $m$-particle effective vertex.
Therefore, vector bosons will be crucial in the construction of $M_{\mathrm{higher}}$. In this subsection, we assume that the vector bosons in $M_{\mathrm{higher}}$ are massless to derive the decomposition relation. Since the contraction of the massless fermion current with massless vector propagator are the same as with massive vector bosons, the decomposition relations are valid for the massive vector bosons. In the massive case, the vector bosons in $M_{\mathrm{lower}}$ are still massless, but the vector bosons in $M_{\mathrm{higher}}$ could be massive.For the convenience, we relabel the momenta of $M_{\mathrm{lower}}$ from $\{p_1,\cdots,p_m\}$ to $\{l_1,\dots,l_{m-n};k_1,\dots,k_n\}$, where $k$ corresponds to the momenta of gauge bosons and $l$ corresponds to the others. Now we write $M_{\mathrm{lower}}$ in terms of polarization $\epsilon$ and vertex $\Gamma$,
\begin{equation}
    M_{\mathrm{lower}}(l_1,\dots,l_{m-n};k_1^{h_1},\dots,k_n^{h_n})=\Gamma^{\mu_1\cdots \mu_n}(k_1,\cdots,k_n)\prod_i\epsilon^{h_i}_{\mu_i}(k_i,r_i)~,
\end{equation}
where $h_i$ and $r_i$ are the helicity and reference momentum of gauge boson $i$. Here $\Gamma$ is not the conventional vertex in Feynman diagrams in SM.

The decomposition relation is based on two key points. One point is the BSM vertex $\Gamma$ is multilinear to the momenta of vector bosons,
\begin{equation}
    \Gamma^{\mu_1\cdots \mu_n}(k_1,\dots,k_n)=\sum_{j_1,\dots,j_n} \Gamma^{\mu_1\cdots \mu_n}(q_{1j_1},\dots,q_{nj_n})~,
    \label{eqn:multilinear}
\end{equation}
where
\begin{equation}
    k_i=\sum_{j_i=1}^{n_i} q_{i j_i},\quad n_i = 1,2.
\end{equation}

The other point is that the current $J_\mu$ is proportional to the polarization vector of a photon $\epsilon_\mu$. They can be written in a uniform notation as
\begin{equation} \begin{aligned}
    J^{(n_i)}_\mu(q_{i1}^{h_{i1}},\dots,q_{in_i}^{h_{in_i}})
    &=F^{(n_i)}_{q_{ij_i}}(q_{i1}^{h_{i1}},\dots,q_{in_i}^{h_{in_i}})\epsilon_\mu^{H_{ij_i}}(q_{ij_i},r_{ij_i})~,
    \label{eqn:current}
\end{aligned} \end{equation}
where $n_i$ is the number of external particles and $F$ is a factor. The helicity $H_{ij_i}$ is a function of $h_{ij_i}$. When $n_i=1$, it reduces to polarization $\epsilon_\mu$ and $H_{i1}=h_{i1}$. In this case, the reference momentum $r_{ij_i}$ is arbitrary. When $n_i=2$, it reduces to the current, whose expression will be given in the next subsection.

Now we express $M_{\mathrm{higher}}$ in terms of $\Gamma$ and $J^{(n_i)}$. Combining Eqs.~\eqref{eqn:multilinear} and \eqref{eqn:current}, it reduces to
\begin{equation} \begin{aligned}
    M_{\mathrm{higher}}
    &=\Gamma^{\mu_1\cdots \mu_n}(k_1,\dots,k_n)\prod_i J^{(n_i)}_\mu(q_{i1}^{h_{i1}},\dots,q_{in_i}^{h_{in_i}})\\
    &=\sum_{j_1,\dots,j_n} \Gamma^{\mu_1\cdots \mu_n}(q_{1j_1},\dots,q_{nj_n})\prod_i F^{(n_i)}_{q_{ij_i}}(q_{i1}^{h_{i1}},\dots,q_{in_i}^{h_{in_i}})\epsilon_\mu^{H_{i j_i}}(q_{ij_i},r_{ij_i})\\
    &=\sum_{j_1,\dots,j_n} F^{(n_i)}_{q_{ij_i}}(q_{i1}^{h_{i1}},\dots,q_{in_i}^{h_{in_i}}) \Gamma^{\mu_1\cdots \mu_n}(q_{1j_1},\dots,q_{nj_n}) \prod_i \epsilon_\mu^{H_{i j_i}}(q_{ij_i},r_{ij_i})\\
    &=\sum_{j_1,\dots,j_n} F^{(n_i)}_{q_{ij_i}}(q_{i1}^{h_{i1}},\dots,q_{in_i}^{h_{in_i}}) M_{\mathrm{lower}}(q_{i1}^{H_{ij_1}},\dots,q_{in_i}^{H_{ij_n}})\\
    \label{decomposition}
\end{aligned} \end{equation}
where we ignored the momenta $l$ in $M_{\mathrm{lower}}$. This is the decomposition relations for helicity amplitudes.

Here the multilinear property of the momenta dependence in the vertex $\Gamma$ is crucial for the decomposition of helicity amplitudes. One may worry about the universality of this multilinear vertex which may limit the usage of the decomposition relation. Our argument is that this multilinear momentum dependent vertex is widely appeared in high dimensional couplings, both in SM framework with loops and effective high dimensional operators. As we will see in the next subsection, the momentum of the vector bosons in these vertices comes from the partial derivative of the vector boson in the Lagrangian, which is commonly exists especially in higher dimensional operators.

\subsection{Applications}
Now we want to know what kind of effective interactions will give multilinear vertices. Since $\Gamma$ is multilinear, each momentum of vector boson should be linear in these vertices. It implies that there is no $D_\mu$ in the effective interactions. All momenta come from the field strength tensor $X_{\mu\nu}$.

In dimension-6 operators, there are only three kinds of effective operators $\psi^2 X\varphi$, $X^2\varphi^2$ and $X^3$ fulfilling these conditions. The corresponding multilinear vertices are
\begin{equation} \begin{aligned}
    (\bar{\psi}\gamma^{\mu\nu}\psi)\varphi X_{\mu\nu}&:
    &\Gamma^\mu(k_1)&=[l_1|k_1\gamma^\mu|l_2]+[l_1|\gamma^\mu k_1|l_2],\\
    \varphi^2 X^{\mu\nu}X_{\mu\nu}&: 
    &\Gamma^{\mu\nu}(k_1,k_2)&=k_1^\mu k_2^\nu-g^{\mu\nu}k_1\cdot k_2,\\
    \mathrm{tr}(X^\mu_\nu X^\nu_\rho X^\rho_\mu )&:
    &\Gamma^{\mu\nu\rho}(k_1,k_2,k_3)&=k_1^\nu k_2^\rho k_3^\mu+\cdots.\\
\end{aligned} \end{equation}
The decomposition relation Eq.~\eqref{decomposition} can be applied to these effective interactions.

From now on, we will return to the $HVV$ vertex from Eq.~\eqref{L_int} and use the label $\{p_1,\dots,p_m\}$. This vertex is bilinear to the momenta of vector bosons, which is
\beq
\Gamma^{\mu\nu}(k,k^{\prime})=
-i\frac{4}{v}[c_{VV}~(k^{\nu} k^{\prime\mu}-k\cdot k^{\prime}g^{\mu\nu})+\tilde{c}_{VV}~\epsilon^{\mu\nu\rho\sigma} k_{\rho}k^{\prime}_{\sigma} ]~,
\label{Gamma}
\eeq
where $k$,~$k^{\prime}$ are the momenta of the two vector bosons.
So when $k=p_2+p_3$ or $k^\prime=p_4+p_5$, or both, where $p_i$s are momentum of external legs,
we have
\bea
\nonumber
&&\Gamma^{\mu\nu}(k,k^{\prime})\\
&=&\Gamma^{\mu\nu}(p_2+p_3,k^{\prime})
=\Gamma^{\mu\nu}(p_2,k^{\prime})+\Gamma^{\mu\nu}(p_3,k^{\prime}) \\
&=&\Gamma^{\mu\nu}(p_2+p_3,p_4+p_5)
=\Gamma^{\mu\nu}(p_2,p_4)+\Gamma^{\mu\nu}(p_2,p_5)
+\Gamma^{\mu\nu}(p_3,p_4)+\Gamma^{\mu\nu}(p_3,p_5).
\eea

\begin{figure}[!htbp]
        \includegraphics[width=0.3\textwidth]{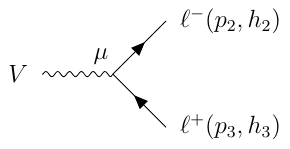}
        \caption{\it \centering The current $J_\mu$ of $V\to\ell^-\ell^+$.}
        \label{fig6}
\end{figure}

On the other hand, we write down the current $J_\mu$ of $V\to\ell^+\ell^-$
in Fig.~\ref{fig6} explicitly,
\bea
J^{(2)}_\mu(p_2^{\mp\frac{1}{2}},p_3^{\pm\frac{1}{2}})
&=&\frac{f^{\mp}_V(s_{23})}{\sqrt{2}}\langle 2^\mp|\gamma_\mu| 3^{\mp}\rangle \\
&=&\pm f^{\mp}_V(s_{23})\langle2^\mp|3^\pm\rangle\epsilon_\mu^\pm(3,2)
\label{J1} \\
&=&\pm f^{\mp}_V(s_{23})\langle2^\pm|3^\mp\rangle\epsilon_\mu^\mp(2,3),
\label{J2}
\eea
where $\epsilon_\mu^\pm(3,2)\equiv\epsilon_\mu^\pm(p_3,p_2)$
could be considered as a polarization vector of photon
with external momentum  $p_3$ and $p_2$ is the chosen reference momentum.
Similarly, $\epsilon_\mu^\pm(2,3)$ represents a photon
with external momentum  $p_2$ with reference momentum  $p_3$.
In principle, $J_\mu$ is a gauge-dependent quantity. As we ignore the mass of leptons,
it could be considered as a gauge-independent quantity in our proof.

According to Eqs.~\eqref{J2} and \eqref{J1}, we know $H_{ij_i}=2h_{ij_i}$ and the expression of $F^{(2)}$ are
\begin{equation} \begin{aligned}
    F^{(2)}_{q_{i1}}(q_{i1}^{-\frac{1}{2}},q_{i2}^{+\frac{1}{2}})
    &=f^-_V(k_i^2)[q_{i1}q_{i2}],\quad
    &F^{(2)}_{q_{i2}}(q_{i1}^{-\frac{1}{2}},q_{i2}^{+\frac{1}{2}})
    &=f^-_V(k_i^2)\langle q_{i1}q_{i2}\rangle,\\
    F^{(2)}_{q_{i1}}(q_{i1}^{+\frac{1}{2}},q_{i2}^{-\frac{1}{2}})
    &=-f^+_V(k_i^2)\langle q_{i1}q_{i2}\rangle,\quad
    &F^{(2)}_{q_{i2}}(q_{i1}^{+\frac{1}{2}},q_{i2}^{-\frac{1}{2}})
    &=-f^+_V(k_i^2)[q_{i1}q_{i2}].\\
\end{aligned} \end{equation}

Based on these equations, we could decompose amplitudes of
$H\to V \gamma\to  \ell \ell \gamma$ as
\bea \nonumber
\mathcal{M}(2^{-}_{\ell^- },3^{+}_{\ell^+},4^{-}_\gamma)
&=&F^{(2)}_{p_2}(p_2^{-\frac{1}{2}},p_3^{+\frac{1}{2}})\mathcal{M}(2_\gamma^-,4_\gamma^-)+F^{(2)}_{p_3}(p_2^{-\frac{1}{2}},p_3^{+\frac{1}{2}})\mathcal{M}(3_\gamma^+,4_\gamma^-)\\ 
&=&f^l_V(s_{23})\times([23]\mathcal{M}(2_\gamma^-,4_\gamma^-)+
\langle 23\rangle\mathcal{M}(3_\gamma^+,4_\gamma^-)),
\label{decomposition_hgamma2l}
\eea

\begin{figure}[!htbp]
        \includegraphics[width=0.8\textwidth]{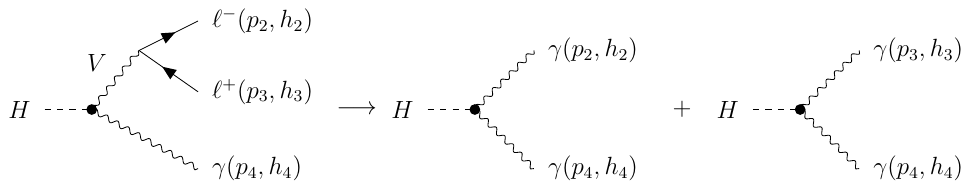}
        \caption{\it \centering Decomposition of amplitudes of $H\to V \gamma\to  \ell\ell\gamma$.}
        \label{fig4}
\end{figure}

In the last step,
the reference momenta of photons are different, which do not affect
the form of $\mathcal{M}(\gamma,\gamma)$ because the vertex $\Gamma^{\mu\nu}$
satisfy Ward identity.
The other helicity amplitudes of $H\to V \gamma\to  \ell \ell \gamma$ have similar decomposition.
An illustrating diagram for Eq.~\eqref{decomposition_hgamma2l} is shown in Fig.~\ref{fig4}.
Each amplitude of $H\to V \gamma\to  \ell\ell\gamma$ is composed by two
amplitudes of $H\to \gamma \gamma$. It degenerate to one term because
the amplitude of $H\to \gamma \gamma$ with reverse helicities is equal to zero. So the $CP$ violation
phase keeps as a global phase in $H\to V \gamma\to  \ell\ell\gamma$ process.

\begin{figure}[!htbp]
        \includegraphics[width=1.0\textwidth]{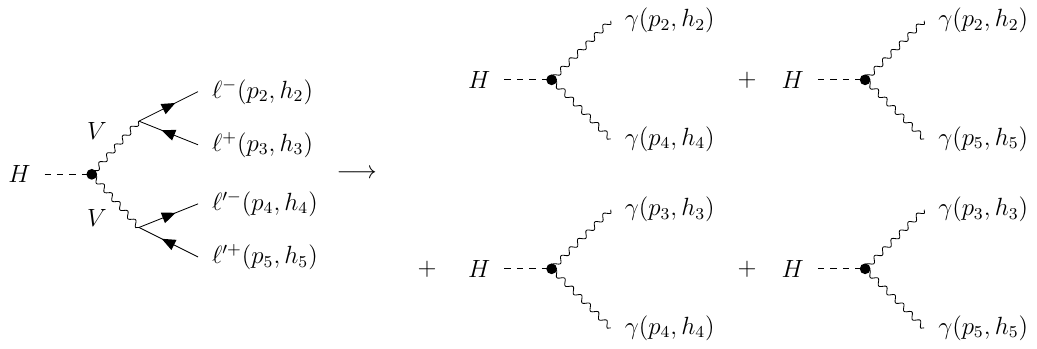}
        \caption{\it \centering Decomposition of amplitudes of $H\to V V \to  4\ell$.}
         \label{fig5}
\end{figure}

Next we prove that decomposition relation is also suitable
for process $H\to V V \to  4\ell $. That is
\bea
\nonumber
&&\mathcal{M}(2^{-}_{\ell^-},3^{+}_{\ell^+ },4^{-}_{\ell^{\prime -}},5^{+}_{\ell^{\prime +}})\\
\nonumber
&=&F^{(2)}_{p_2}(p_2^{-\frac{1}{2}},p_3^{+\frac{1}{2}})F^{(2)}_{p_4}(p_4^{-\frac{1}{2}},p_5^{+\frac{1}{2}})\mathcal{M}(2_\gamma^-,4_\gamma^-)\\ \nonumber
&&+F^{(2)}_{p_2}(p_2^{-\frac{1}{2}},p_3^{+\frac{1}{2}})F^{(2)}_{p_5}(p_4^{-\frac{1}{2}},p_5^{+\frac{1}{2}})\mathcal{M}(2_\gamma^-,5_\gamma^+)\\ \nonumber
&&+F^{(2)}_{p_3}(p_2^{-\frac{1}{2}},p_3^{+\frac{1}{2}})F^{(2)}_{p_4}(p_4^{-\frac{1}{2}},p_5^{+\frac{1}{2}})\mathcal{M}(3_\gamma^+,4_\gamma^-)\\
&&+F^{(2)}_{p_3}(p_2^{-\frac{1}{2}},p_3^{+\frac{1}{2}})F^{(2)}_{p_5}(p_4^{-\frac{1}{2}},p_5^{+\frac{1}{2}})\mathcal{M}(3_\gamma^+,5_\gamma^+)
\label{eq:h4l-1}\\
\nonumber
&=& f^l_V(s_{23})f^l_V(s_{45})\times (\\
\nonumber
 &&\quad   [23][45]\mathcal{M}(2_\gamma^-,4_\gamma^-)
\nonumber
+ [23]\langle 45\rangle\mathcal{M}(2_\gamma^-,5_\gamma^+)\\
 &&+ \langle 23\rangle [45] \mathcal{M}(3_\gamma^+,4_\gamma^-)
 + \langle 23\rangle\langle 45\rangle \mathcal{M}(3_\gamma^+,5_\gamma^+)~),
\label{eq:h4l-2}
\eea
where for the specific helicity states in Eq.~\eqref{eq:h4l-1}, one term of $H\to 4\ell$ is  decomposed into four terms. 
Furthermore, the four terms in Eq.~\eqref{eq:h4l-2} would degenerate to two terms since reverse-sign $H\to\gamma\gamma$ amplitudes are zero, as shown in Eq.~(\ref{eqn:hzz}). The illustrating diagrams are shown in Fig.~\ref{fig5}.

One may think it's ridiculous with the first glance at the decomposition of the scattering amplitude of $H\to 4\ell$ into the combination of four $H\to \gamma\gamma$ amplitudes, as the final decay products of leptons are changed strangely to be photons.  We argue that this result is instructive and has profound physical meanings. The amplitude of the Higgs decay can be considered as a function of the momentum of its decay products. Here we just find the form of the momentum dependence between the $H\to 4\ell$ and  the $H\to \gamma\gamma$ amplitudes.  One can easily generate this results into similar processes obeying the two aforementioned  key preconditions. Our results provide a new viewpoint of the amplitude of the multiple decay of the Higgs.

\subsection{$CP$ violation phase in helicity amplitudes}
From the decomposition relations, we see that the amplitudes of $H\to\gamma\gamma$
are basis for other amplitudes. Since $\mathcal{M}(+,-)=\mathcal{M}(-,+)=0$,
the left basis are $\mathcal{M}(+,+)$ and $\mathcal{M}(-,-)$.
$CP$ violation phases are reverse in the two bases.
In $H\to\gamma\gamma$ and $H\to V\gamma \to \ell\ell\gamma$ processes,
$CP$ violation phase is a global phase in each amplitude.
So generally speaking it is an unobservable phase if one doesn't consider
interference between this amplitude and the background amplitudes~\cite{Chen:2017plj,Wan:2017qiq}.
In $H\to 4\ell$ process, two bases coexist in each amplitude,
thus the $CP$ violation phase appear as a physical observable.
Meanwhile, it means that the interference between $CP$-even term
and $CP$-odd term exists at differential cross section level
after squaring the amplitude. So the interference could be probed
through kinematic angles~\cite{He:2019kgh,CMS:2019ekd,Chatrchyan:2013mxa,Anderson:2013afp}.
An obvious effect is a shift of azimuthal angle
caused by the interference
between $CP$-even and $CP$-odd term. So we see the $CP$ phase angle dependence clearly in $HVV$ processes as the benefit of our amplitude decomposition relations.


\section{BSM amplitudes from on-shell approach (massless)\label{section:on-shell}}

In on-shell approach, the amplitude is not derived from Lagrangian and
Feynman rules. Instead, it is constructed directly from on-shell particle states.
In this section, firstly we introduce spinor variables for particles,
secondly
we show how amplitudes of $H\gamma\gamma$ are represented
 and fixed, thirdly we get amplitudes of $H\gamma\ell\ell$ and $H 4\ell$
through recursion relations.

\subsection{Spinor variables}
The right-handed and left-handed spinors in Eq.~\eqref{eq:epsilon}
have their two-component versions~\cite{Dixon:1996wi,Dixon:2013uaa}:
\bea
| i_\alpha \rangle \equiv \lambda_{i\alpha}\equiv u_+(p_i) \equiv |i^+ \rangle,&\quad&
|i^{\dot\alpha} ] \equiv
\tilde\lambda^{\dot\alpha}_i\equiv u_-(p_i) \equiv |i^- \rangle,\nonumber \\
\langle i^\alpha | \equiv \lambda_{i}^\alpha\equiv \overline{u_-(p_i)} \equiv \langle i^- |,
&\quad&
[i_{\dot\alpha} | \equiv
\tilde\lambda_{i\dot\alpha}\equiv \overline{u_+(p_i)} \equiv \langle i^+ |,
\eea
where the spinor indices can be raised or lowered by antisymmetric tensors
$\epsilon^{\alpha\beta}$ and $\epsilon_{\alpha\beta}$ ,
\beq
\lambda^\alpha=\epsilon^{\alpha\beta}\lambda_\beta,\quad
\lambda_\alpha=\epsilon_{\alpha\beta}\lambda^\beta~.
\eeq
In this notation,
\beq
\langle i j \rangle \equiv \lambda_{i}^\alpha \lambda_{j\alpha}, \quad
[ij] \equiv \tilde\lambda_{i\dot\alpha} \tilde\lambda^{\dot\alpha}_j~.
\eeq
An on-shell momentum of a massless particle is represented as
\beq
p_{\alpha\dot{\alpha}} \equiv p_\mu\sigma^{\mu}_{\alpha\dot{\alpha}}
=\lambda_\alpha\tilde{\lambda}_{\dot{\alpha}},
\eeq
where $\sigma^{\mu}=(1,\vec{\sigma})$ with $\vec{\sigma}$ being the Pauli matrices.

\subsection{Amplitude of $H\gamma\gamma$}

\begin{figure}[!htbp]
        \includegraphics[width=0.8\textwidth]{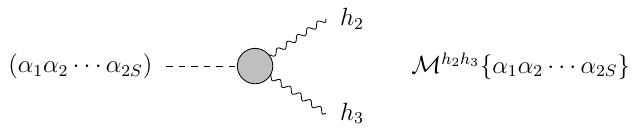}
        \caption{\it \centering A general one massive and two massless particle interaction.
The subscript S represents spin of the massive particle, $h_2$, $h_3$ are helicities of two massless particles. }
        \label{diagram_hgammagamma}
\end{figure}

A general three point amplitude with one massive
and two massless particle interaction is shown in Fig.~\ref{diagram_hgammagamma}~\cite{Arkani-Hamed:2017jhn}, where $\alpha_i$, $i=1,2,...,2S$ are indices of spinors,
$S$ represents spin of the massive particle, $h_2$, $h_3$ are helicities of the two massless particles.

For amplitude of $H\gamma\gamma$,
as the massive particle $H$ is a scalar with zero spin,
 we don't need to
take care about its spinors, which makes the formula much simpler.
A general ansatz is~\cite{Arkani-Hamed:2017jhn,Shadmi:2018xan,Dixon:2004za}
\beq
\mathcal{M}_3(1_H,2_\gamma^{h_2},3_\gamma^{h_3})=e^{i\xi^{h_2,h_3}}\frac{g}{m^{h_2+h_3-1}}
[23]^{h_2+h_3}~,
\eeq
where $\xi^{h_2,h_3}$ represents a helicity-related phase,
$g$ represents an overall coupling constant, $m$ is the mass of the Higgs boson.
As $\langle 23\rangle [32]=(p_2+p_3)^2=p_1^2=m^2$,~$\langle 23\rangle =\frac{m^2}{[32]}$.
The little group scaling~\cite{Elvang:2015rqa,Cheung:2017pzi} requires $h_2+h_3=2h_2=2h_3$, so $\mathcal{M}(2_\gamma^{+},3_\gamma^{-})=
\mathcal{M}(2_\gamma^{-},3_\gamma^{+})=0$. The non-zero amplitudes are only
$\mathcal{M}(2_\gamma^{+},3_\gamma^{+})$  and $\mathcal{M}(2_\gamma^{-},3_\gamma^{-})$.
It doen't lose generality to require $\xi^{+,+}=-\xi^{-,-}=\xi^\prime$
 since their equal part could be absorbed into the redefinition of $g$.

 The inequality of $|\mathcal{M}(2_\gamma^{+},3_\gamma^{+})|\neq|\mathcal{M}(2_\gamma^{-},3_\gamma^{-})|$
could also cause $CP$ violation, however, this is 
not favored by physics assumption. Specifically,
from Lagrangian in Eq.~\eqref{L_int} we need 
$C_{VV}$ and $\tilde{C}_{VV}$ to be real to keep 
the Lagrangian Hermitian conjugate, so 
it results in $|\mathcal{M}(2_\gamma^{+},3_\gamma^{+})|=|\mathcal{M}(2_\gamma^{-},3_\gamma^{-})|$  as in Eq.~\eqref{eqn:hgammagamma}. From the above discussion, the nonzero amplitudes are
\bea
\mathcal{M}_3(1_H,2_\gamma^{+},3_\gamma^{+})&=&e^{i\xi^\prime}\frac{g}{m}
[23]^2~, \\
\mathcal{M}_3(1_H,2_\gamma^{-},3_\gamma^{-})&=&e^{-i\xi^\prime}\frac{g}{m}
\langle 23 \rangle^2~,
\label{eqn:onshell-hgammagamma}
\eea
which is equal to Eq.~\eqref{eqn:hgammagamma} as long as
we require $\frac{g}{m}=\frac{2c^S_{\gamma\gamma}}{v}$ and
$\xi^\prime=\xi$.

\subsection{Amplitudes of $H\to \gamma \ell \ell$}

\begin{figure}[!htbp]
        \includegraphics[width=1.0\textwidth]{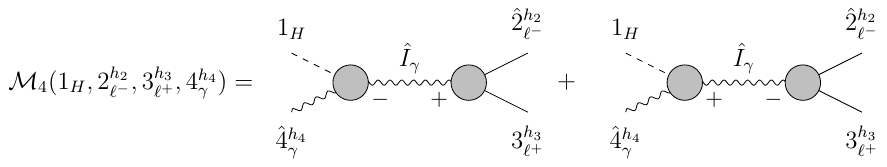}
        \caption{\it \centering Factorization of $H\to \gamma \ell \ell$.
We take the mediate particle as $\gamma$ for simplicity.
}
        \label{diagram_hgammall}
\end{figure}

The amplitudes of $H\to \gamma \ell \ell$
could be built from three point amplitudes by recursion relations.
For the amplitude of $H\to \gamma \ell \ell$, a factorization
way is $H\to \gamma V, V\to \ell \ell$.
Figure \ref{diagram_hgammall} shows this factorization.
The mediate particle is taken as $\gamma$ to aviod
amplitude of massive particles, its momentum is marked as
``$I$".
We shift momenta of the $2,4$ external particles according to
BCFW recursion relation approach~\cite{Britto:2004ap,Britto:2005fq,Feng:2011np}.
That is
\beq
|\hat{2}]=|2],
\quad |\hat{4}]=|4]+z|2],
\quad|\hat{4}\rangle=|4\rangle,
\quad |\hat{2}\rangle=|2\rangle-z|4\rangle,
\label{eq:24shift}
\eeq
where $z$ is a complex number, and shifted momenta are hatted.

The corresponding analytical formula are
\bea
\mathcal{M}(1_{H},2^{h_2}_{\ell^- },3^{h_3}_{\ell^+},4^{h_4}_\gamma)
=&&P_\gamma(s_{23})
\mathcal{M}(1_H,\hat{4}^{h_4}_\gamma,-\hat{P}^{-}_{I\gamma})
\mathcal{M}(\hat{P}^{+}_{I\gamma},\hat{2}^{h_2}_{\ell^- },3^{h_3}_{\ell^+})~\nonumber\\
&+&P_\gamma(s_{23})
\mathcal{M}(1_H,\hat{4}^{h_4}_\gamma,-\hat{P}^{+}_{I\gamma})
\mathcal{M}(\hat{P}^{-}_{I\gamma},\hat{2}^{h_2}_{\ell^- },3^{h_3}_{\ell^+})~,
\label{eq:onshell-hgammall}
\eea
where
$\hat{p}_I=p_1+\hat{p}_4=-(\hat{p}_2+p_3$) is momentum of the mediate photon,
$P_\gamma(s_{23})=1/s_{23}=1/(p_2+p_3)^2$ is the propagator with unshifted momenta.

The helicity amplitudes of $\gamma\ell^-\ell^+$ are three point amplitudes
with massless particles,
which are fully fixed by little group scaling and dimension analysis,
\bea
\label{eq:onshell-gammall_1}
\mathcal{M}({1}^{-}_\gamma,{2}^{-}_{\ell^- },3^{+}_{\ell^+})
&=&\tilde{e}\frac{\langle 1 2\rangle^2}{\langle 23\rangle}~,  \\
\label{eq:onshell-gammall_2}
\mathcal{M}({1}^{-}_\gamma,{2}^{+}_{\ell^- },3^{-}_{\ell^+})
&=&\tilde{e}\frac{\langle 1 3\rangle^2}{\langle 23\rangle}~,  \\
\label{eq:onshell-gammall_3}
\mathcal{M}({1}^{+}_\gamma,{2}^{-}_{\ell^- },3^{+}_{\ell^+})
&=&\tilde{e}\frac{[ 1 3]^2}{[23]}~,  \\
\label{eq:onshell-gammall_4}
\mathcal{M}({1}^{+}_\gamma,{2}^{+}_{\ell^- },3^{-}_{\ell^+})
&=&\tilde{e}\frac{[ 1 2]^2}{[23]}~,
\eea
where $\tilde{e}=-\sqrt{2}e$~, Eqs.~\eqref{eq:onshell-gammall_1}\eqref{eq:onshell-gammall_2}
correspond to $[23]=0$ solution and Eqs.~\eqref{eq:onshell-gammall_3}\eqref{eq:onshell-gammall_4}
correspond to $\langle 23 \rangle=0$ solution.

After inserting Eq.~\eqref{eqn:onshell-hgammagamma} and Eq.~\eqref{eq:onshell-gammall_1} into
Eq.~\eqref{eq:onshell-hgammall}, we get
\bea
\mathcal{M}(1_{H},2^{-}_{\ell^- },3^{+}_{\ell^+},4^{-}_\gamma)
&=&\tilde{e}P_\gamma(s_{23})\times\frac{2c^S_{\gamma V}}{v}e^{-i\xi}
\frac{\langle \hat{I} \hat{4} \rangle^2
[\hat{I}3]^2}{[\hat{2}3]}, \nonumber\\
&=&\tilde{e}P_\gamma(s_{23})\times\frac{2c^S_{\gamma V}}{v}e^{-i\xi}
[23]\langle 24\rangle^2,
\label{eq:onshell-hgammall-mpm}
\eea
where the last equation is because
\beq
\langle \hat{4}\hat{I} \rangle [\hat{I} 3]=\langle \hat{4}|\hat{p}_I |3]
=\langle \hat{4}|\hat{p}_2+p_3 |3]
=\langle \hat{4}|\hat{p}_2 |3]
=\langle \hat{4} \hat{2} \rangle [\hat{2} 3]
=\langle 42 \rangle [23]~,
\label{eq:4223}
\eeq
and meanwhile an analytical continuum of $|-p\rangle = -|p\rangle$,
$|-p] = |p]$ is adopted. 
So Eq.~\eqref{eq:onshell-hgammall-mpm} is the same formula as
the one derived in effective Lagrangian calculation~(see Eq.~\eqref{eqn:hzgamma}).
It is worthy to notice that because $P_\gamma(s_{23})=\frac{1}{\langle 23 \rangle [32]}$,
Eq.~\eqref{eq:onshell-hgammall-mpm} is
proportional to  $\frac{\langle 24\rangle^2}{\langle 23 \rangle}$
and thus has a singularity when $\langle 23 \rangle=0$.

If we take the propagator $V$ as a $Z$ boson,
we should consider a $H\gamma Z$ amplitude together with a $Z\gamma\gamma$ amplitude.
The $H\gamma Z$ amplitude is an amplitude with two massive one massless particles,
and the $Z\gamma\gamma$ amplitude is an amplitude with one massive two massless particles.
They are more complex than the $H\gamma\gamma$ amplitude since the spin of $Z$ is $1$.
These two amplitudes should use bolded spinor variables ~\cite{Arkani-Hamed:2017jhn,Shadmi:2018xan}.


\subsection{Amplitudes of $H\to 4\ell$}
\begin{figure}[!htbp]
        \includegraphics[width=1.0\textwidth]{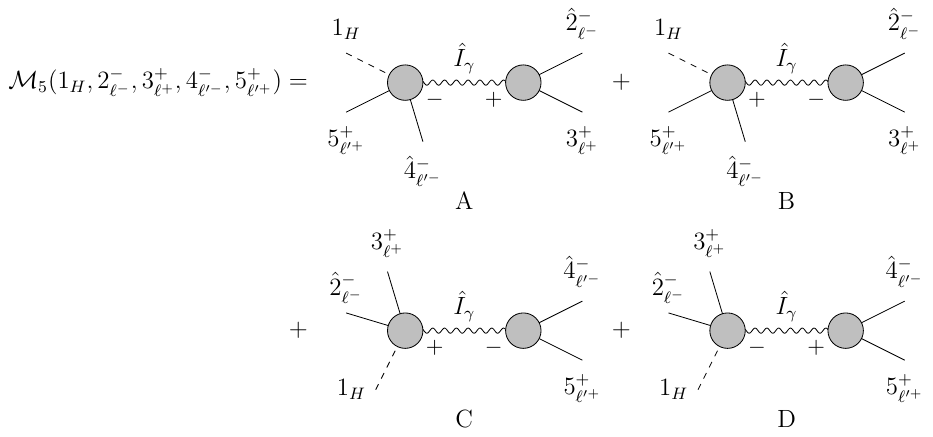}
        \caption{\it \centering Factorization of $H\to 4\ell$. The external legs
are arranged in clockwise order.
}
        \label{diagram_h4l}
\end{figure}

The amplitude of $H\to 4\ell$ is a five point amplitude, we could
factorize it into two parts: a four point amplitude plus a
three point amplitude.
Each amplitude split into
four parts as shown in Fig.~\ref{diagram_h4l}.

In formula, it is
\bea
\mathcal{M}_5(1_{H},2^{-}_{\ell^- },3^{+}_{\ell^+},4^{-}_{\ell^{\prime-} },
5^{+}_{\ell^{\prime+}})
=&&P_\gamma(s_{23})\mathcal{M}(1_H,\hat{4}^{-}_{\ell^{\prime-}},5^{+}_{\ell^{\prime+}},-\hat{P}^{-}_{I\gamma})
\mathcal{M}(\hat{P}^{+}_{I\gamma},\hat{2}^{-}_{\ell^- },3^{+}_{\ell^+})\nonumber \\
&+&P_\gamma(s_{23})\mathcal{M}(1_H,\hat{4}^{-}_{\ell^{\prime-}},5^{+}_{\ell^{\prime+}},-\hat{P}^{+}_{I\gamma})
\mathcal{M}(\hat{P}^{-}_{I\gamma},\hat{2}^{-}_{\ell^- },3^{+}_{\ell^+})\nonumber \\
&+& P_\gamma(s_{45})\mathcal{M}(1_H,\hat{2}^{-}_{\ell^- },3^{+}_{\ell^+},-\hat{P}^{+}_{I\gamma})
\mathcal{M}(\hat{P}^{-}_{I\gamma},\hat{4}^{-}_{\ell^{\prime-}},5^{+}_{\ell^{\prime+}})
\nonumber \\
&+& P_\gamma(s_{45})\mathcal{M}(1_H,\hat{2}^{-}_{\ell^- },3^{+}_{\ell^+},-\hat{P}^{-}_{I\gamma})
\mathcal{M}(\hat{P}^{+}_{I\gamma},\hat{4}^{-}_{\ell^{\prime-}},5^{+}_{\ell^{\prime+}})
,
\label{eq:onshell-h4l-mpmp}
\eea
which corresponds to diagram A, B, C, D respectively.
Diagram A and B correspond to $(1,4,5)+(2,3)$ factorization,
Diagram C and D correspond to  $(1,2,3)+(4,5)$ factorization.
We assumed $\ell\neq \ell^\prime$ for generality, so
the factorizations of $(1,2,5)+(3,4)$ and $(1,3,4)+(2,5)$ are absent because of flavor symmetry.
Next we calculate these four diagrams separately.

The formula for diagram A is
\bea
&&P_\gamma(s_{23})
\mathcal{M}(1_H,\hat{4}^{-}_{\ell^{\prime-}},5^{+}_{\ell^{\prime+}},-\hat{P}^{-}_{I\gamma})
\mathcal{M}(\hat{P}^{+}_{I\gamma},\hat{2}^{-}_{\ell^- },3^{+}_{\ell^+})
\nonumber\\ \label{eq:onshell-h4l-mpmp-1-1}
&=&\frac{2c^S_{\gamma \gamma}}{v}e^{-i\xi}P_\gamma(s_{23}) P_\gamma(s_{\hat{4}5})[\hat{4}5]\langle \hat{4}\hat{I} \rangle^2
\frac{[\hat{I}3]^2}{[\hat{2}3]} \nonumber \\
\label{eq:onshell-h4l-mpmp-1-2}
&=& \frac{2c^S_{\gamma \gamma}}{v}e^{-i\xi}
P_\gamma(s_{23}) P_\gamma(s_{45}) [45][23]\langle 24\rangle^2~,
\label{eqn:m5-1}
\eea
where
in the last step we have used
\beq
\langle \hat{4}\hat{I} \rangle[\hat{I} 3]=\langle \hat{4}|\hat{p}_2+p_3|3]
=\langle \hat{4}|\hat{p}_2|3]=\langle \hat{4}\hat{2} \rangle[\hat{2} 3]
=\langle 42 \rangle[23]
\eeq
as in Eq.~\eqref{eq:4223} and
\beq
P_\gamma(s_{\hat{4}5})[\hat{4}5]=\frac{-1}{\langle \hat{4}5 \rangle[\hat{4}5]} [\hat{4}5]
=\frac{-1}{\langle 45 \rangle[45]} [45]
=P_\gamma(s_{45})[45]~,
\eeq
$\langle \hat{2}3\rangle=0$ is chosen for three-point amplitude,
which is also required in diagram B.


The formula for diagram B is
\bea
&&P_\gamma(s_{23})
\mathcal{M}(1_H,\hat{4}^{-}_{\ell^{\prime-}},5^{+}_{\ell^{\prime+}},-\hat{P}^{+}_{I\gamma})
\mathcal{M}(\hat{P}^{-}_{I\gamma},\hat{2}^{-}_{\ell^- },3^{+}_{\ell^+})
\nonumber \\
&=&\frac{2c^S_{\gamma \gamma}}{v}e^{i\xi}P_\gamma(s_{23})
P_\gamma(s_{\hat{4}5})\langle\hat{4}5\rangle[5\hat{I}]^2
\times\frac{\langle\hat{I}\hat{2}\rangle^2}{\langle\hat{2}3\rangle}\nonumber \\
&=&\frac{2c^S_{\gamma \gamma}}{v}e^{i\xi} P_\gamma(s_{23})P_\gamma(s_{\hat{4}5})\langle\hat{4}5\rangle[5\hat{I}]^2\times 0 \nonumber \\
&=& 0,
\label{eqn:m5-2}
\eea
where the 3-point amplitude is equal
 to zero because $\langle\hat{2}3\rangle=\langle\hat{I}\hat{2}\rangle=0$~.

The formula for diagram C is
\bea
&&P_\gamma(s_{45})\mathcal{M}(1_H,\hat{2}^{-}_{\ell^- },3^{+}_{\ell^+},-\hat{P}^{+}_{I\gamma})
\mathcal{M}(\hat{P}^{-}_{I\gamma},\hat{4}^{-}_{\ell^{\prime-}},5^{+}_{\ell^{\prime+}})
\nonumber  \\ \label{eq:4part}
&=&\frac{2c^S_{\gamma \gamma}}{v}e^{i\xi}
P_\gamma(s_{45})
P_\gamma(s_{\hat{2}3})
\langle \hat{2}3\rangle [3\hat{I}]^2
\times\frac{\langle \hat{I}\hat{4}\rangle^2}{\langle \hat{4}5\rangle}\nonumber \\
&=&\frac{2c^S_{\gamma \gamma}}{v}e^{i\xi}
P_\gamma(s_{45})
P_\gamma(s_{23}) \langle 23\rangle \langle 45\rangle [35]^2~,
\label{eqn:m5-3}
\eea
where
\beq
[3\hat{I}]\langle \hat{I}\hat{4}\rangle=[35]\langle 5\hat{4}\rangle
\eeq
and
\beq
P_\gamma(s_{\hat{2}3})\langle \hat{2}3\rangle=\frac{-1}{\langle \hat{2}3 \rangle[\hat{2}3]} \langle \hat{2}3 \rangle
=\frac{-1}{\langle 23 \rangle[23]} \langle 23\rangle
=P_\gamma(s_{23}) \langle 23\rangle~
\eeq
are used. $[\hat{4}5]=0$ is required.

The formula for diagram D is
\bea
&&P_\gamma(s_{45})\mathcal{M}(1_H,\hat{2}^{-}_{\ell^- },3^{+}_{\ell^+},-\hat{P}^{-}_{I\gamma})
\mathcal{M}(\hat{P}^{+}_{I\gamma},\hat{4}^{-}_{\ell^{\prime-}},5^{+}_{\ell^{\prime+}})
\nonumber \\ \label{eq:3part}
&=&\frac{2c^S_{\gamma \gamma}}{v}e^{-i\xi}
P_\gamma(s_{45})
P_\gamma(s_{\hat{2}3})
[\hat{2}3]\langle\hat{2}\hat{I}\rangle^2
\frac{[\hat{I}5]^2}{[\hat{4}5]}\nonumber \\
&=&0~,
\label{eqn:m5-4}
\eea
where $[\hat{4}5]=[\hat{I}5]=0$ makes the three point amplitude zero.


After summing up the results of four parts, that is adding
Eq.s~\eqref{eqn:m5-1}\eqref{eqn:m5-2}\eqref{eqn:m5-3}\eqref{eqn:m5-4} together,
we get
\bea
\mathcal{M}(1_{H},2^{-}_{\ell^- },3^{+}_{\ell^+},4^{-}_{\ell^{\prime-} },
5^{+}_{\ell^{\prime+}})
&=&
\quad \frac{2c^S_{\gamma \gamma}}{v}e^{-i\xi}
P_\gamma(s_{23}) P_\gamma(s_{45}) [45][23]\langle 24\rangle^2
\\
&&+\frac{2c^S_{\gamma \gamma}}{v}e^{i\xi}
P_\gamma(s_{45})
P_\gamma(s_{23}) \langle 23\rangle \langle 45\rangle [35]^2~,
\eea
which has the same form as the one derived in effective Lagrangian calculation~(see Eq.~\eqref{eqn:hzz}).
So we get a consistent result from the on shell approach.
Boundary contributions here are supposed to be zero.

\section{BSM amplitudes from on-shell approach (massive)\label{section:on-shellmassive}}

When the propagators are $Z$ bosons or $W$ bosons, the massless on-shell method 
won't work and we use the little-group covariant massive spinor formalism~\cite{Arkani-Hamed:2017jhn,Durieux:2019eor,Wu:2021nmq}.  In this section, we deduce $HVV$ amplitudes according to the presumed $HVV$ vertex. A more general method starting from an ansatz is left in Appendix~\ref{appendixA}.
\subsection{Massive spinor formalism}
In massive spinor formalism~\cite{Arkani-Hamed:2017jhn}, a massive momentum is 
decomposed into two light-like vectors and thus two pairs of massless spinors,
\beq
\bs{p}_{\alpha\dot{\alpha}}=\lambda^I_\alpha\tilde{\lambda}_{I\dot{\alpha}}
=|\bs{p}^I\rangle[\bs{p}_I|,\quad \mbox{and} \quad 
\bar{\bs{p}}^{\dot{\alpha}\alpha}=-\tilde\lambda^{I\dot\alpha}\lambda_I^{\alpha}
=-|\bs{p}^I]\la\bs{p}_I|~.
\label{eq:MassiveSpinor}
\eeq  
Where $I=1,2$ is little-group index, $p$ is bolded to denote massive momentum.
The equation of motion reads,
\beq
\bs{p}|\bs{p}^I]= m |\bs{p}^I\ra,\quad
\bar{\bs{p}}|\bs{p}^I\ra= m |\bs{p}^I],\quad
[\bs{p}^I|\bar{\bs{p}}=-m\la \bs{p}^I|,\quad
\la \bs{p}^I| \bs{p} =-m [\bs{p}^I|.
\label{eq:eom}
\eeq
The polarized vector of a massive vector boson of momentum $\bs{p}$ and mass $m$
is~\cite{Chung:2018kqs}
\beq
\epsilon_\mu^{IJ}(\bs{p})=\frac{1}{\sqrt{2}m}\la\bs{p}^I|\gamma_\mu|\bs{p}^J],
\eeq 
which corresponds to two transverse and one longitudinal modes,
\beq
\epsilon^+_\mu\equiv \epsilon^{11}_\mu,\quad 
\epsilon^0_\mu \equiv \frac{1}{2}(\epsilon^{12}_\mu+\epsilon^{21}_\mu),\quad
\epsilon^-_\mu\equiv \epsilon^{22}_\mu~.
\label{eq:epsilon-massive}
\eeq

\subsection{Amplitudes of $HVV$ }
\label{massiveHVV}
According to the $HVV$ vertex in Eq.~\eqref{Gamma}, the amplitude of $HVV$ is
\beq
\mathcal{M}(\bs{1}_H,\bs{I}_V,\bs{J}_V)=\Gamma^{\mu\nu}(\bs{p}_I,\bs{p}_J)
\epsilon_\mu(\bs{p}_I)\epsilon_\nu(\bs{p}_J)
\eeq
Since $p_I^{\mu} p_J^{\nu} \epsilon_\mu(p_I)\epsilon_\nu(p_J)=0$, we could add this term in amplitude formula freely and make amplitude more symmetric.
\bea
\mathcal{M}(\bs{1}_H,\bs{I}_V,\bs{J}_V)&=&
-\frac{4}{v}[c_{VV}(\bs{p}_I^{\mu} \bs{p}_J^{\nu}+
\bs{p}_I^{\nu} \bs{p}_J^{\mu}-\bs{p}_I\cdot \bs{p}_Jg^{\mu\nu})+\tilde{c}_{VV}\epsilon^{\mu\nu\alpha\beta} \bs{p}_{I\alpha}\bs{p}_{J\beta} ]\epsilon_\mu(\bs{p}_I)\epsilon_\nu(\bs{p}_J)
\nonumber \\
&=&-\frac{1}{v}[c_{VV}{\rm tr}(\gamma^\mu\gamma^\alpha\gamma^\nu\gamma^\beta)+
i\tilde{c}_{VV}{\rm tr}(\gamma^\mu\gamma^\nu\gamma^\alpha\gamma^\beta\gamma^5)]\bs{p}_{I\alpha}\bs{p}_{J\beta}
\epsilon_\mu(\bs{p}_I)\epsilon_\nu(\bs{p}_J)
 \nonumber \\
&=&-\frac{1}{v}c^S_{VV}[e^{-i\xi}{\rm tr}(\sigma^\mu\bar{\sigma}^\alpha\sigma^\nu
\bar{\sigma}^\beta)+
e^{i\xi}{\rm tr}(\bar{\sigma}^\mu{\sigma}^\alpha\bar{\sigma}^\nu
\sigma^\beta)]
\bs{p}_{I\alpha}\bs{p}_{J\beta}
\epsilon_\mu(\bs{p}_I)\epsilon_\nu(\bs{p}_J)~,
\label{eq:mhvv}
\eea
where 

\bea
&&4(g^{\mu\alpha}g^{\nu\beta}-g^{\mu\nu}g^{\alpha\beta}+g^{\mu\beta}g^{\nu\alpha})={\rm tr}(\gamma^\mu\gamma^\alpha\gamma^\nu\gamma^\beta)=
{\rm tr}(\sigma^\mu\bar{\sigma}^\alpha\sigma^\nu
\bar{\sigma}^\beta)+{\rm tr}(\bar{\sigma}^\mu{\sigma}^\alpha\bar{\sigma}^\nu
\sigma^\beta),\\ 
&&-4i\epsilon^{\mu\nu\alpha\beta}=
{\rm tr}(\gamma^\mu\gamma^\nu\gamma^\alpha\gamma^\beta\gamma^5)
=-{\rm tr}(\gamma^\mu\gamma^\alpha\gamma^\nu\gamma^\beta\gamma^5)=
-{\rm tr}(\sigma^\mu\bar{\sigma}^\alpha\sigma^\nu
\bar{\sigma}^\beta)+{\rm tr}(\bar{\sigma}^\mu{\sigma}^\alpha\bar{\sigma}^\nu
\sigma^\beta)
\eea
are used. Eq.~\eqref{eq:mhvv} shows a general formula for 
$HVV$ amplitudes, which constitutes two parts with opposite 
$CP$ violation phases. If one part is zero, the $CP$ violation
phase degenerates to a trivial phase. 
After inserting Eq.~\eqref{eq:eom}\eqref{eq:epsilon-massive}
 into Eq.~\eqref{eq:mhvv},
we get 
\beq
\mathcal{M}(\bs{1}_H,\bs{I}_V,\bs{J}_V)=
\frac{2c^S_{VV}}{v}[e^{-i\xi}\la \bs{IJ}\ra ^2+e^{i\xi} [ \bs{IJ}] ^2]~.
\eeq
When one vector boson is a massless photon, the amplitudes become
\bea
\mathcal{M}(\bs{1}_H,I^+_\gamma,\bs{J}_V)&=&
\frac{2c^S_{\gamma V}}{v}e^{i\xi} [ I\bs{J}] ^2~,\nonumber  \\ 
\mathcal{M}(\bs{1}_H,I^-_\gamma,\bs{J}_V)&=&
\frac{2c^S_{\gamma V}}{v}e^{-i\xi} \la I\bs{J}\ra ^2~,
\eea
where the compact form for a general amplitude decompose into 
two parts, each has a trivial $CP$ violation phase. 
When both of the two vector bosons are massless photons, the amplitudes become
\bea
\mathcal{M}(\bs{1}_H,I^+_\gamma,J^+_\gamma)&=&
\frac{2c^S_{\gamma \gamma}}{v}e^{i\xi} [ IJ] ^2~,\nonumber  \\
\mathcal{M}(\bs{1}_H,I^+_\gamma,J^-_\gamma)&=&
0 ~,\nonumber \\
\mathcal{M}(\bs{1}_H,I^-_\gamma,J^+_\gamma)&=&
0~, \nonumber \\
\mathcal{M}(\bs{1}_H,I^-_\gamma,J^-_\gamma)&=&
\frac{2c^S_{\gamma \gamma}}{v}e^{i\xi} \la IJ \ra ^2~.\nonumber  \\
\eea
So the general amplitude decompose into four parts, each with 
a trivial global $CP$ violation phase except for the zero ones.

The amplitude of $Vl^- l^+$ with massless leptons is
\bea
\mathcal{M}(\bs{I}_V,2^-_{l^-},3^+_{l^+})&=&
el_V\la 2|\gamma^\mu|3]\epsilon_\mu(\bs{p}_I) 
=\sqrt{2}e l_V\frac{\la 2\bs{I}\ra [3 \bs{I}]}{m_V}, \\
\mathcal{M}(\bs{I}_V,2^+_{l^-},3^-_{l^+})&=&
e r_V[ 2|\gamma^\mu|3\ra \epsilon_\mu(\bs{p}_I) 
=
\sqrt{2} e r_V\frac{[ 2\bs{I}] 
\la 3 \bs{I}\ra }{m_V},\\
\mathcal{M}(\bs{I}_V,2^-_{l^-},3^-_{l^+})&=&
\mathcal{M}(\bs{I}_V,2^+_{l^-},3^+_{l^+})
=0.
\eea

\subsection{Amplitudes of $H\to 4\ell$ with massive propagators}

\begin{figure}[!htbp]
        \includegraphics[width=0.5\textwidth]{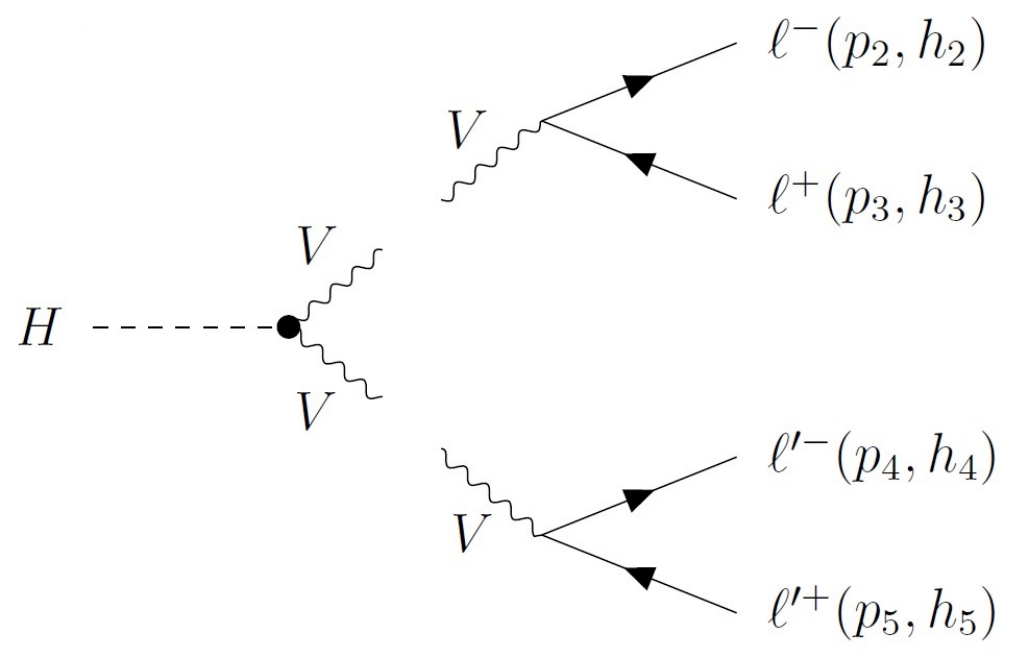}
        \caption{\it \centering Glue amplitudes by contracting little-group indices of the massive propagators.}
        \label{4V}
\end{figure}

We get the amplitude of $H\to VV \to 4\ell$ by simply gluing 
the amplitudes of $H\to VV$, $V\to \ell^+\ell^-$ and 
$V\to \ell^{\prime+}\ell^{\prime-}$ as shown in Fig.~\ref{4V}.

When a propagator goes on-shell, the amplitude factorize into the tensor product of two subamplitudes.
\begin{equation}
    \lim_{p^2\rightarrow m^2}\mathcal{M}=\frac{\mathcal{M}_L^{\{I_1... I_{2s}\}}\otimes
   \mathcal{M}_R^{\{J_1... J_{2s}\}}}{p^2-m^2},
\end{equation}
where $L,R$ represent left and right amplitudes for each gluing. 
For each propagator particle, the sign of its momentum is opposite 
in the left and the right amplitudes, as shown explicitly in 
Eq.~\eqref{eq:onshell-hgammall}. An analytical continuum 
$|-p\rangle = -|p\rangle$, $|-p \rb = |p \rb $ is adopted. The gluing procedure is done by choosing the singlet of the little-group for the on-shell propagator 
\begin{equation}
    \mathcal{M}_L^{\{I_1... I_{2s}\}}\otimes
   \mathcal{M}_R^{\{J_1... J_{2s}\}}=\mathcal{M}_{L,\{I_1... I_{2s}\}}\epsilon^{I_1J_1}...\epsilon^{I_{2s}J_{2s}}
   \mathcal{M}_{R,\{J_1... J_{2s}\}}.
   \label{eq:glue}
\end{equation}

Since amplitude of $H\to VV \to 4\ell$ has two propagators $p_I$ and $p_J$, we take the limit $p_I^2\rightarrow m^2_V$ and $p_J^2\rightarrow m^2_V$ simultaneously,
\begin{equation} \begin{aligned}
    &\lim_{p_I^2,p_J^2\rightarrow m^2_V} \mathcal{M}(1_{H},2^{-}_{\ell^- },3^{+}_{\ell^+},4^{-}_{\ell^{\prime-} },5^{+}_{\ell^{\prime+}})\\
    =&f^-_V(s_{23}) f^-_V(s_{45})\mathcal{M}(\bs{I}_V,2^-_{\ell^- },3^+_{\ell^+})\otimes\mathcal{M}(\bs{1}_H,\bs{I}_V,\bs{J}_V)\otimes\mathcal{M}(\bs{I}_V,4^-_{\ell^{\prime-} },5^+_{\ell^{\prime+}})\\
    =&\frac{2c^S_{VV}}{v}f^-_V(s_{23}) f^-_V(s_{45})\left[e^{-i\xi}\lim_{p_I^2,p_J^2\rightarrow m^2_V}\mathcal{M}^a+e^{i\xi}\lim_{p_I^2,p_J^2\rightarrow m^2_V}\mathcal{M}^b\right],\\
\end{aligned} \end{equation}
where $\mathcal{M}^a$ and $\mathcal{M}^b$ in this limit are
\begin{equation} \begin{aligned}
    &\lim_{p_I^2,p_J^2\rightarrow m^2_V}2\mathcal{M}^a\\
    =&\langle 2I^{I_1}\rangle\langle I_{I_1}J_{J_1}\rangle\langle J^{J_1}4\rangle[3I^{I_2}]\langle I_{I_2}J_{J_2}\rangle[J^{J_2}5]+\langle 2I^{I_1}\rangle\langle I_{I_1}J_{J_1}\rangle[J^{J_2}5][3I^{I_2}]\langle I_{I_2}J_{J_2}\rangle\langle J^{J_2}4\rangle\\
    =&-\sqrt{p_I^2}\sqrt{p_J^2}\left(\langle 24\rangle[3|p_I p_J|5]+\langle 2|p_J|5]\langle3|p_I|4]\right),\\
\end{aligned} \end{equation}

\begin{equation} \begin{aligned}
    &\lim_{p_I^2,p_J^2\rightarrow m^2_V}2\mathcal{M}^b\\
    =&\langle 2I^{I_1}\rangle[I_{I_1}J_{J_1}]\langle J^{J_1}4\rangle[3I^{I_2}][ I_{I_2}J_{J_2}][J^{J_2}5]+\langle 2I^{I_1}\rangle[I_{I_1}J_{J_1}][J^{J_2}5][3I^{I_2}][I_{I_2}J_{J_2}]\langle J^{J_2}4\rangle\\
    =&-\sqrt{p_I^2}\sqrt{p_J^2}\left(\langle 2|p_I p_J|4\rangle[35]+\langle 2|p_I|5]\langle3|p_J|4]\right),\\
\end{aligned} \end{equation}
where we used
\beq
|\bs{p}^I\ra_\alpha\la\bs{p}_I|^\beta=-m\delta^{~~\beta}_\alpha,\quad
|\bs{p}^I]^{\dot\alpha}[\bs{p}_I|_{\dot\beta}=m\delta^{\dot\alpha}_{~~\dot\beta}.
\eeq

When the propagators go off-shell, we should use
\begin{equation} \begin{aligned}
    p_I=p_2+p_3,\quad p_J=p_4+p_5,\quad \sqrt{p_I^2}=\sqrt{p_J^2}=m_V,\\
\end{aligned} \end{equation}
instead of $p_I$ and $p_J$. Now we get
\begin{equation} \begin{aligned}
    \mathcal{M}^a=m_V^2[45][23]\langle 24\rangle^2,\quad 
    \mathcal{M}^b=m_V^2\langle 23\rangle \langle 45\rangle [35]^2.
\end{aligned} \end{equation}
So the amplitude of $H\to VV \to 4\ell$ is
\beq
    \mathcal{M}(1_{H},2^{-}_{\ell^- },3^{+}_{\ell^+},4^{-}_{\ell^{\prime-} },
    5^{+}_{\ell^{\prime+}})=
    \frac{2c^S_{VV}}{v}
    f^-_V(s_{23}) f^-_V(s_{45})
    [e^{-i\xi}[45][23]\langle 24\rangle^2+
    e^{i\xi}\langle 23\rangle \langle 45\rangle [35]^2],
    \label{eq:4lMassive}
\eeq
 which is same as the former results in Eq.~\eqref{eqn:hzz}.

In Eq.~\eqref{eq:glue},
the particles are all on-shell especially for the propagator particles.
In our specific process, 
before and after gluing, $s_{23}=s_{45}=m_V^2$ should be required
in the amplitude except for the propagator factor.
So Eq.~\eqref{eq:4lMassive} is the amplitude of  
$H\to VV \to 4\ell$ in the on-shell limit. 
By contrast, Eq.~\eqref{eqn:hzz} is obtained from off-shell method 
and $s_{23}\neq   s_{45}\neq m_V^2$ . 
Why do these two amplitudes are same? 
The reason is because  
the amplitude of $H\to VV \to 4\ell$
is independent of the residue $z$. 
Eq.~\eqref{eq:4223} illustrate this point explicitly, 
the combined amplitude has no $z$ dependence.

\subsection{ $CP$ violation phase }

In the on-shell way we get a compact form to show that the 
$CP$ violation phase in $HVV$ amplitude
is not a trivial phase because of the 
BSM $HVV$ vertex. 
It degenerates to a trivial phase once 
the helicity of the vector boson is fixed as in $H\gamma\gamma$
and $H\to \gamma V$ cases.
By contrast, in the off-shell way, we only see the 
nontriviality of the $CP$ violation phase in $H\to 4\ell$ amplitude.
It is because in the off-shell way we don't deal with 
massive vector boson independently. 
Its full properties are exhibited indirectly in the four final states.

\section{Summary and discussion\label{section:summary}}

The $HVV$ amplitudes with $CP$ violation in beyond Standard Model are analyzed in two ways.
One way is the off-shell way under field theory framework, we decompose helicity amplitudes
of $H\to \gamma V \to \gamma \ell\ell $ and $H\to VV  \to 4\ell$ into helicity amplitudes of
$H\to\gamma\gamma$. 
 There are two preconditions for the decomposition relation. 1.The multilinear momentum dependence of the $HVV$ vertexes, which allow us to decompose the vertexes of the overall momentum into a summation of momentum of sub-processes. 2. The current of $J_\mu$ in $V\to\ell^+\ell^-$ is  formally proportional to a photon's polarization vector, which allow us to replace such a sub-process by an equivalent photon. 
The other way is through the on-shell scattering amplitude approach.
For the massless propagator case, the 3-point amplitude of $H\gamma\gamma$ is the start point,
then the 4-point amplitude of $H \ell\ell \gamma$ and
 the 5-point amplitude $H 4\ell$ are obtained through
recursion relations.
For the massive propagator case, we adopt the little-group covariant 
massive-spinor formalism.
It expresses $HVV$ amplitude firstly, then glue  $V\ell\ell$ amplitudes 
to get final $H\to VV  \to 4\ell$ amplitudes. 
We get consistent results through off-shell and on-shell ways.

The $CP$ violation phase in $H\to VV  \to 4\ell$ amplitude is
a nontrivial phase while in $H\to \gamma V \to \gamma \ell\ell $ and
$H\to\gamma\gamma$ amplitudes it is a global trivial phase.
In off-shell way, the decomposition relations shows that in $H\to VV  \to 4\ell$ amplitude, it mix
 the $H\to\gamma\gamma$ amplitudes with different helicities,
so it mix $H\to\gamma\gamma$ amplitudes with different dependences
 on $CP$ violation phases.
In the on-shell way, the nontriviality of the $CP$ violation phase
appears directly in $HVV$ amplitudes, which is not a helicity amplitude
but a compact massive-spinor amplitude.
It degenerates to a trivial phase when
the helicity of at least one vector boson is fixed, 
 such as that happens in $H\gamma V$ and $H\gamma\gamma$ amplitudes.
The decays of $V\to \ell \ell$ maintains the nontriviality
of the $CP$ violation phase.
The on-shell way supplies a simpler and clearer viewpoint
about the $CP$ violation phase in amplitudes. Our systematic analysis on the series of the amplitudes of $HVV$ processes exhibits the dependence of $CP$ violation phase, therefore, can be convenient for the future $CP$ violation searches in $HVV$ couplings.

\begin{acknowledgements}
We are grateful to Chih-Hao Fu, Kang Zhou for useful discussion about BCFW recursion relations.
X.W. thanks Bo Feng for Scattering Amplitude School 2021 in Hangzhou, China.
The work is supported by the National Natural Science Foundation of China under Grant No.~11847168, the Fundamental Research Funds for the Central Universities of China under Grant No.~GK202003018, and the Natural Science Foundation of Shannxi Province, China (2019JM-431, 2019JQ-739).

\end{acknowledgements}

\begin{appendix}

\section{massive $HVV$ amplitudes}\label{appendixA}
\setcounter{equation}{0}
\renewcommand{\theequation}{A.\arabic{equation}}

In section \ref{massiveHVV}, we derived the massive $HVV$ amplitudes by using the $HVV$ vertex $\Gamma^{\mu\nu}$. Now we construct them directly. Consider the three massive amplitude $\mathcal{M}(\bs{1}_H,\bs{I}_V,\bs{J}_V)$. In this case, Ref.~\cite{Arkani-Hamed:2017jhn} showed that the spinor space is spanned by two tensors, the symmetric tensor $\mathcal{O}_{\beta\gamma}$ and the antisymmetric tensor $\varepsilon_{\beta\gamma}$. We choose the first tensor
\begin{equation} \begin{aligned}
    \mathcal{O}_{\beta\gamma}=p_{2\{\beta\dot{\gamma}} p^{\dot{\gamma}}_{3\gamma\}}=|2_J\rangle_{\{\beta}[2^J 3^K]\langle 3_K|_{\gamma\}}+(\beta\leftrightarrow\gamma).\\
\end{aligned} \end{equation}
Therefore, the three massive amplitude have a general form
\begin{equation} \begin{aligned}
    \mathcal{M}(\bs{1}_H,\bs{I}_V,\bs{J}_V)&=\lambda_2^{\beta_1 J_1}\lambda_2^{\beta_2 J_2}\lambda_3^{\gamma_1 K_1}\lambda_3^{\gamma_2 K_2}\sum_{i=0}^1 g_{\sigma_i}(\mathcal{O}^{2-i}\varepsilon^i)_{\{\beta_1\beta_2\},\{\gamma_1\gamma_2\}}\\
    &=\lambda_2^{\beta_1 J_1}\lambda_2^{\beta_2 J_2}\lambda_3^{\gamma_1 K_1}\lambda_3^{\gamma_2 K_2}\left(g_{\sigma_0}(\mathcal{O}\mathcal{O})_{\{\beta_1\beta_2\},\{\gamma_1\gamma_2\}}+g_{\sigma_1}(\mathcal{O}\varepsilon)_{\{\beta_1\beta_2\},\{\gamma_1\gamma_2\}}\right).\\
    \label{eqn:HVVexpand}
\end{aligned} \end{equation}
The second term $(\mathcal{O}\varepsilon)$ is
\begin{equation} \begin{aligned}
    \mathcal{O}_{\beta_1\gamma_1}\varepsilon_{\beta_2\gamma_2}+\mathcal{O}_{\beta_1\gamma_2}\varepsilon_{\beta_2\gamma_1}
    \rightarrow \langle\mathbf{2}\mathbf{3}\rangle[\mathbf{2}\mathbf{3}].
\end{aligned} \end{equation}
Since this term is symmetric between angle and square brackets, it doesn't contribute to the $CP$ violation. The first term $(\mathcal{O}\mathcal{O})$ can be parametrized as
\begin{equation} \begin{aligned}
    &g_1\mathcal{O}_{\beta_1\beta_2}\mathcal{O}_{\gamma_1\gamma_2}+g_2(\mathcal{O}_{\beta_1\gamma_1}\mathcal{O}_{\beta_2\gamma_2}+\mathcal{O}_{\beta_1\gamma_2}\mathcal{O}_{\beta_2\gamma_1})\\
    =&\frac{g_1}{2}(2\mathcal{O}_{\beta_1\beta_2}\mathcal{O}_{\gamma_1\gamma_2}-\mathcal{O}_{\beta_1\gamma_1}\mathcal{O}_{\beta_2\gamma_2}-\mathcal{O}_{\beta_1\gamma_2}\mathcal{O}_{\beta_2\gamma_1})+\frac{2g_2+g_1}{2}(\mathcal{O}_{\beta_1\gamma_1}\mathcal{O}_{\beta_2\gamma_2}+\mathcal{O}_{\beta_1\gamma_2}\mathcal{O}_{\beta_2\gamma_1})\\
    =&\frac{g_1}{2}m^4_V(\varepsilon_{\beta_1\gamma_1}\varepsilon_{\beta_2\gamma_2}+\varepsilon_{\beta_1\gamma_2}\varepsilon_{\beta_2\gamma_1})+\frac{2g_2+g_1}{2}(\mathcal{O}_{\beta_1\gamma_1}\mathcal{O}_{\beta_2\gamma_2}+\mathcal{O}_{\beta_1\gamma_2}\mathcal{O}_{\beta_2\gamma_1}),\\
\end{aligned} \end{equation}
where we use Schouten identity
\begin{equation} \begin{aligned}
    &2\mathcal{O}_{\beta_1\beta_2}\mathcal{O}_{\gamma_1\gamma_2}-\mathcal{O}_{\beta_1\gamma_1}\mathcal{O}_{\beta_2\gamma_2}-\mathcal{O}_{\beta_1\gamma_2}\mathcal{O}_{\beta_2\gamma_1}\\
    =&\lambda^{J_1}_{2\{\beta_1}\lambda^{K_1}_{3\beta_2\}}\lambda^{J_2}_{2\{\gamma_1}\lambda^{K_2}_{3\gamma_2\}}([2_{J_1} 3_{K_1}][2_{J_2} 3_{K_2}]-[2_{J_1} 3_{K_2}][2_{J_2} 3_{K_1}])\\
    =&\lambda^{J_1}_{2\{\beta_1}\lambda^{K_1}_{3\beta_2\}}\lambda^{J_2}_{2\{\gamma_1}\lambda^{K_2}_{3\gamma_2\}}[2_{J_1} 2_{J_2}][3_{K_1}3_{K_2}]\\
    =&m^2_V\lambda^{J_1}_{2\{\beta_1}\lambda^{K_1}_{3\beta_2\}}\lambda^{J_2}_{2\{\gamma_1}\lambda^{K_2}_{3\gamma_2\}}\varepsilon_{J_1 J_2}\varepsilon_{K_1 K_2}\\
    =&m^4_V(\varepsilon_{\beta_1 \gamma_1}\varepsilon_{\beta_2 \gamma_2}+\varepsilon_{\beta_1 \gamma_2}\varepsilon_{\beta_2 \gamma_1}).\\
\end{aligned} \end{equation}

Therefore, the $(\mathcal{O}\mathcal{O})$ term give two independent structure that contribute to the $CP$ violation,
\begin{equation} \begin{aligned}
    \mathcal{O}_{\beta_1\gamma_1}\mathcal{O}_{\beta_2\gamma_2}+\mathcal{O}_{\beta_1\gamma_2}\mathcal{O}_{\beta_2\gamma_1}&\rightarrow&[\mathbf{2}\mathbf{3}]^2,\\
    \varepsilon_{\beta_1\gamma_1}\varepsilon_{\beta_2\gamma_2}+\varepsilon_{\beta_1\gamma_2}\varepsilon_{\beta_2\gamma_1}&\rightarrow&\langle\mathbf{2}\mathbf{3}\rangle^2.\\
\end{aligned} \end{equation}

\end{appendix}

\bibliographystyle{utphys} 
\bibliography{reference}

\end{document}